\documentclass[usenatbib,referee]{biom}
\def\bSig\mathbf{\Sigma}

\newcommand{\nn}{\nonumber}
\newcommand{\Ind}{1\!\mathrm{l}}
\usepackage{graphicx,comment}
\usepackage{amsfonts,amsmath,amssymb,bbm,float}

\usepackage{natbib,xcolor}

\title[Addressing patient heterogeneity in disease studies]{Addressing patient heterogeneity in disease predictive model development}

\author
{Xu Gao$^1$, Weining Shen$^1$, Jing Ning$^2$, Ziding Feng$^3$,  Jianhua Hu$^{4,*}$\email{jh3992@cumc.columbia.edu}  \\
 $^{1}$ Department of Statistics, University of California, Irvine, CA, USA
  \\
$^{2}$ Department of Biostatistics, The University of Texas MD Anderson Cancer Center, Houston, TX, USA \\
$^{3}$  Cancer Prevention Program, Fred Hutchinson Cancer Research Center, Seattle, WA, USA 
 \\
 $^{4}$ Department of Biostatistics, Columbia University, New York, NY, USA
}
 
\begin{document}

\date{{\it Received October} 2007. {\it Revised February} 2008.  {\it
Accepted March} 2008.}

\pagerange{\pageref{firstpage}--\pageref{lastpage}} 
\volume{64}
\pubyear{2008}
\artmonth{December}


\doi{10.1111/j.1541-0420.2005.00454.x}


\label{firstpage}


\begin{abstract}
This paper addresses patient heterogeneity associated with prediction problems in biomedical applications. We propose a systematic hypothesis testing approach to determine the existence of patient subgroup structure and the number of subgroups in patient population if subgroups exist. A mixture of generalized linear models is considered to model the relationship between the disease outcome and patient characteristics and clinical factors, including targeted biomarker profiles. We construct a test statistic based on expectation maximization (EM) algorithm and derive its asymptotic distribution under the null hypothesis. An important computational advantage of the test is that the involved parameter estimates under the complex alternative hypothesis can be obtained through a small number of EM iterations, rather than optimizing the objective function. We demonstrate the finite sample performance of the proposed test in terms of type-I error rate and power, using extensive simulation studies. The applicability of the proposed method is illustrated through an application to a multi-center prostate cancer study.  \\
\end{abstract}

%



\maketitle

\section{Introduction}
Large heterogeneity is known to exist in many diseases, including cancers. It is not surprising that inhomogeneity among patients is observed in various characteristics, including basic health information, genomic profiles, and the treatment effects. Hence, it is critical to address disease heterogeneity among patients for improving the accuracy of disease diagnosis and personalized treatment. 
Ignoring or mistreating the heterogeneity issue  can lead to poor biological and clinical conclusions in biomarker evaluations for early detection, diagnosis, prognosis, and treatment of a disease.  However, statistical tools that account for potentially hidden subgroups are lacking, and most studies today are performed by either ignoring heterogeneity or relying on ad hoc methods to account for heterogeneity.  Our particular interest is to build formal predictive models for clinical outcome variables of interest based on biomarkers and other clinical factors, an essential step in translating biological discovery to clinical applications, while appropriately accounting for likely patient heterogeneity.

Several approaches have been proposed in the literature concerning identification of the latent subgroups based on observed characteristics (e.g., biomarkers) and study of the differences in their effects among subgroups, including the methods under the finite mixture model framework  \citep{Dayton1988,Lanza2013}, tree-based method  \citep{Loh2015}, and recursive partitioning method \citep{Lip2011,Lip2017}. Bayesian approaches have also gained popularity in this area due to their flexibility in conducting inference. For example, \citet{Berger2014}  accounts for multiplicity in testing statistics among subgroups and provides individual probabilities of treatment effect. A Bayesian shrinkage estimation method was introduced to select and estimate the sparse subgroup effects \citep{Jones2011, Ratkovic2017}. 

The aforementioned methods together with other related work \citep{Bonetti:2004, Cai:2011, Foster:2011, Song:2004,Zhao2013} generally construct subgroups under the assumption of the existence of subgroups, which tends to assign patients to subgroups even when no meaningful subgroups exist. Another common drawback of these methods is their incapability of formally controlling false positives. For this purpose, off-the-shelf model selection criteria, such as Bayesian information criterion (BIC), have been  used mainly because of their convenient implementation but often without careful justifications. Recent attempts have been made to develop confirmatory tests for the existence of the subgroups and for characterizing the subgroup effects. For example, \citet{Shen2015} considered a structured logistic-normal mixture model and proposed a statistical test for the existence of subgroups. Its extension to address heterogeneous subgroup variances using a penalized likelihood approach is discussed by \citet{Shen2017}. However, this line of work is so far limited to consideration of at most two subgroups and it remains unclear how the proposed methodology can be extended to allow multiple subgroups. Additional work in this area includes a change plane approach proposed by \citet{Fan2017}, and a penalized regression approach by \citet{Ma2017}. 

Some important issues are unresolved by the prior mixture-model-based methods. First, the existing work focuses on testing the existence of subgroups, without further rigorous testing procedures on the number of subgroups. Second, most of the existing methods pose restrictive model assumptions on the predictors and the response, e.g., a logistic model for binary outcomes and a Cox model for survival outcomes. To solve those issues, we consider modeling the relationship between the response and the predictors using a mixture of generalized linear models (GLMs) and propose a hypothesis testing procedure to determine the number of subgroups with differential biomarker/clinical covariate effects on the response variable. Examples include a structured logistic mixture model where different subgroups are allowed to have differential covariate effects on the binary response probability (e.g., cancer or not). This model-based approach also reduces false positives in subgroup analysis, and enables us to evaluate variables that are predictive of subgroup membership with probabilistic interpretations. Our method is motivated by the idea of expectation-maximization test on the order of a finite normal mixture model proposed in a series of work  \citep{Chen:2009,Li2010,chen:2012}, due to its attraction of the typically required small number of iterations in implementation without the need of obtaining the optimal solution for the objective function. However, the main drawback of most existing work is that a mixture of normal location (or location-scale) model has been focused on, without allowing for the existence of covariates. In addition, there are more challenges in both computation and theoretical development as we consider the more flexible GLM framework. For example, deriving the asymptotic distribution for the test statistic under the null hypothesis becomes much more complicated, and computationally demanding as the number of parameters increases; and our main contribution in this paper is to fill this gap. 

 
 \vspace{-0.2in}
 
\section{Method}
\subsection{The framework}
We consider a GLM framework, where $y$ is a scalar clinical outcome, $\bm{x} \in \mathbb{R}^p$ denotes the $p$ covariates that may have distinct subgroup effects on the outcome; and  $\bm{z} \in \mathbb{R}^q$ denotes the $q$ covariates that share the same effect across different subgroups. Let $f(\bm{y},\bm{x},\bm{z};\bm{\theta},\bm{\gamma})$ be the density function from the GLM, where $\bm{\theta} \in  \mathbb{R}^p$, $\bm{\gamma} \in  \mathbb{R}^q$ are regression coefficients for $\bm{x}$ and $\bm{z}$, respectively. We consider a finite mixture of GLMs, denoted by 
\begin{align}\label{model1}
\sum_{h=1}^m \alpha_h f(y,\bm{x},\bm{z}; \bm{\theta}_h, \bm{\gamma}), 
\end{align}
where $m$ is the number of mixtures, $\alpha_1,\ldots,\alpha_m$ are non-negative weights that sum up to one, $\bm{\theta}_i = (\theta_{i1},\ldots,\theta_{ip})^T$ represents the covariate effect of $\bm{x}$ in subgroup $i$ for $i=1,\ldots,m$, and $\bm{\gamma}$ is the common covariate effect of $\bm{z}$ in the overall population. 

Model \eqref{model1} is widely applicable in practice. For example, we may choose $y$ to be a binary response (e.g., status of disease occurrence), $\bm{x}$ to be a set of biomarkers that can likely help identify the sub-population at higher risk of disease development, and $\bm{z}$ to be a set of baseline covariates. The density form $f$ allows different ways to characterize the relationship between the response and covariates.  

For convenience, we define a mixing distribution
\begin{align}
\Psi(\bm{\theta}) = \sum_{h=1}^m \alpha_h \text{I}(\theta_{h1} \leq \theta_1,\ldots,\theta_{hp} \leq \theta_p),~~\text{for any}~\bm{\theta} = (\theta_1,\ldots,\theta_p)^T \in \mathbb{R}^p.
\end{align}
Then the mixture density function in \eqref{model1} can be written as
\begin{align}
f(y,\bm{x},\bm{z};\Psi,\bm{\gamma})  = \int f(y,\bm{x},\bm{z};\bm{\theta},\bm{\gamma}) d\Psi(\bm{\theta}).
\end{align}
We propose a sequential testing procedure, at each step of which we aim to test
\begin{equation} \label{testlog}
H_0: m=m_0 \mbox{ versus } H_a: m> m_0, \mbox{ where } m_0=1, 2,\cdots.
\end{equation}
We start with the test on $m_0=1$, and move to the second test on $m_0=2$ if $m_0=1$ is rejected; we continue the test until a specific $m_0=m^*$ cannot be rejected, and declare $m^*$ as the possible number of subgroups.  An advantage of this conservative sequential test is that the false positive rate can be controlled at a nominal level due to the nested structure of the sequential null hypotheses. In addition to statistical testing, we also investigate practical factors, such as the sample size, to facilitate determining biologically meaningful subgroups.  We propose to conduct each test in \eqref{testlog} based on the idea of EM test in \citet{Li2010}. However \cite{Li2010} only deals with the simplest case without covariates, and thus is not directly applicable to our case of structured mixture models that are more complicated in terms of both theory and implementation.  
To construct a convenient test statistic,  we adopt $m=2m_0$ as the working alternative hypothesis $H_a$. This working alternative enables us to utilize a likelihood-ratio type test in a computationally and theoretically convenient way.  In the simpler case without covariates, it has been shown in \citet{Li2010} that the power of the test remains solid when in truth the number of subgroups is not $2m_0$.

\subsection{EM test}
Next, we introduce the EM test for the testing problem $H_0: m=m_0$  versus $H_a: m = 2 m_0.$ Given independent and identically distributed (i.i.d) observations $(y_1,\bm{x}_1,\bm{z}_1),\ldots,(y_n,\bm{x}_n,\bm{z}_n)$ and under $H_0$, we consider the log-likelihood function
\begin{align}
l_n(\Psi;\bm{\gamma}) = \sum_{i=1}^n \log \{f(y_i,\bm{x}_i,\bm{z}_i;\Psi,\bm{\gamma})\}.
\end{align}
By maximizing $l_n(\Psi;\bm{\gamma})$, we obtain the maximizer of $\Psi$ and $\bm{\gamma}$, denoted by 
\begin{align}
\hat{\Psi}_0 =\sum_{h=1}^{m_0} \hat{\alpha}_{0h} \Ind(  \hat{\bm{\theta}}_{0h1} \leq \theta_1,\ldots, \hat{\bm{\theta}}_{0hp} \leq \theta_p),~\text{and}~ \hat{\bm{\gamma}}_0,
\end{align} 
where $\hat{\bm{\theta}}_{0h} = (\hat{\theta}_{0h1},\ldots,\hat{\theta}_{0hp})^T$ for every $h=1,\ldots,m_0$.

We then move on to consider the alternative hypothesis $H_a$. The key idea is to obtain a restricted maximum likelihood estimator for the parameters where the mixture weights are not allowed being too close to $0$. This is accomplished by splitting each of the original subgroup $h$ under $H_0$ into two further subgroups, each with weights $\beta_h \alpha_h$ and $(1-\beta_h) \alpha_h$, where $\beta_h \in (0, 0.5]$ is a tuning parameter for every $h=1,\ldots,m_0$. We will discuss the choice of values for $\beta_h$ later. Once the weights are determined, we proceed with the usual EM approach to update the estimated regression coefficients for each of the $2m_0$ subgroups. More specifically, let $\bm{1}$ be a column vector of $1$'s with dimension $p$, and $\eta_h = ( \hat{\bm{\theta}}_{0h}^T \bm{1} +  \hat{\bm{\theta}}_{0 h+1}^T \bm{1})/2$. Consider a partition of $\mathbb{R}$, $\{I_1,\ldots,I_{m_0-1}\}$ with $I_h = (\eta_{h-1},\eta_h]$ for $h=1,\ldots,m_0$. For every fixed $\bm{\beta}=(\beta_1,\ldots,\beta_{m_0})^T \in (0,0.5]^{m_0}$, we create a class of mixing distributions of order $2m_0$:
\begin{align*}
&\Omega_{2m_0}(\bm{\beta}) = \Big\{ \sum_{h=1}^{m_0} \alpha_h \beta_h \Ind(\theta_{1h1} \leq \theta_1,\ldots,\theta_{1hp} \leq \theta_p,\ldots ) +
\alpha_h (1 - \beta_h) \\
& ~~~~\Ind(\theta_{2h1} \leq \theta_1,\ldots,\theta_{2hp} \leq \theta_p,\ldots ): \bm{\theta}_{1h}^T \bm{1}, \bm{\theta}_{2h}^T \bm{1} \in I_h,h=1,\ldots,m_0 \Big\}.
\end{align*}
For each $\Psi \in \Omega_{2m_0}(\bm{\beta})$, we consider a penalized log-likelihood function by including an additional terms in $l_n$, 
\begin{align}
pl_n (\Psi,\bm{\gamma},\bm{\beta}) = l_n(\Psi,\bm{\gamma}) + p(\bm{\beta}), 
\end{align}
in which $p(\bm{\beta}) = \sum_{i=1}^{m_0} p(\beta_i)$ and $p(\beta)$ is a continuous increasing penalty function that satisfies $p(\beta) \leq 0$ and $p(.5)=0$. \textcolor{blue}{In this paper, we use the same penalty function as the one in \citet{Li2010} for convenience, that is, $p(\beta) = C \log (1 - |1 - 2\beta|)$ for some tuning parameter $C > 0$. We will discuss how to choose the value of $C$ in the simulation section. 
 }

We now proceed to the description of the EM algorithm. Let $\mathcal{B}$ be a collection of numbers in $(0,0.5]$, e.g., $\mathcal{B}=\{0.1,0.3,0.5\}$. For each $\bm{\beta}_0 \in \mathcal{B}^{m_0}$, we compute
\begin{align}
\left\{\Psi^{(1)}(\bm{\beta}_0),\bm{\gamma}^{(1)}(\bm{\beta}_0)\right\} = \text{argmax}_{\Psi,\bm{\gamma}} \{ pl_n(\Psi,\bm{\gamma}): \Psi \in \Omega_{2m_0}(\bm{\beta}_0)\}, \nn
\end{align}
with respect to $\bm{\alpha}=(\alpha_1,\ldots,\alpha_{m_0})^T$, and two matrices $\bm{\theta}_1 = (\bm{\theta}_{11},\ldots,\bm{\theta}_{1m_0})$, $\bm{\theta}_2 = (\bm{\theta}_{21},\ldots,\bm{\theta}_{2m_0})$. 
The EM iteration proceeds as follows. Let $\bm{\beta}^{(1)} = \bm{\beta}_0$. Suppose we already obtained $\Psi^{(k)}(\bm{\beta}_0)$ with $\bm{\alpha}^{(k)}$, $\bm{\theta}_1^{(k)}$,$\bm{\theta}_2^{(k)}$, $\bm{\beta}^{(k)}$ and $\bm{\gamma}^{(k)}$ at step $k$. Then we update their values at step $(k+1)$. Define
\begin{align*}
w_{i1h}^{(k)} = \frac{\alpha_{h}^{(k)} \beta_h^{(k)} f(Y_i,\bm{X}_i,\bm{Z}_i; \bm{\theta}_{1h}^{(k)},\bm{\gamma}^{(k)})}{f(Y_i,\bm{X}_i,\bm{Z}_i; \Psi^{(k)}(\bm{\beta}_0),\bm{\gamma}^{(k)})}~~\text{and}~~
w_{i2h}^{(k)} = \frac{\alpha_{h}^{(k)} (1 - \beta_h^{(k)}) f(Y_i,\bm{X}_i,\bm{Z}_i; \bm{\theta}_{2h}^{(k)},\bm{\gamma}^{(k)})}{f(Y_i,\bm{X}_i,\bm{Z}_i; \Psi^{(k)}(\bm{\beta}_0),\bm{\gamma}^{(k)})}
\end{align*}
for $h=1,\ldots,m_0$ and $i=1,\ldots,n$. Then we obtain $\Psi^{(k+1)}(\bm{\beta}_0)$ by calculating
\begin{align}
& \alpha_h^{(k+1)} = n^{-1} \sum_{i=1}^n \left(w_{i1h}^{(k)} + w_{i2h}^{(k)}\right), \\
& \bm{\theta}_{jh}^{(k+1)} = \text{argmax}\left\{ \sum_{i=1}^n w_{ijh}^{(k)} \log f(Y_i,\bm{X}_i,\bm{Z}_i; \bm{\theta},\bm{\gamma}^{(k)}) - n q_{\lambda}(\bm{\theta})\right\} ,~j=1,2 \label{eq232}\\
& \beta_h^{(k+1)} = \text{argmax}\left\{\sum_{i=1}^n w_{i1h}^{(k)} \log(\beta) + \sum_{i=1}^n w_{i2h}^{(k)} \log(1-\beta) + p(\beta)  \right\}.
\end{align}
Meanwhile, $\bm{\gamma}$ is updated by
\begin{align}
\bm{\gamma}^{(k+1)} =  \text{argmax}\left\{ \sum_{i=1}^n w_{ijh}^{(k+1)} \log f(Y_i,\bm{X}_i,\bm{Z}_i; \bm{\theta}^{(k+1)},\bm{\gamma})\right\}.
\end{align}
The above computation will be repeated for $K$ times. For each $\bm{\beta}_0 \in \mathcal{B}^{m_0}$ and $k$, define
\begin{align}
M_n^{(k)}(\bm{\beta}_0) = 2\{pl_n(\Psi^{(k)},\bm{\gamma}^{(k)},\bm{\beta}_0 ) - pl_n^0(\hat{\Psi}_0,\hat{\bm{\gamma}})\}, 
\end{align}
where $pl_n^0$ is the log-likelihood function under $H_0$ without the penalization term, and $\hat{\Psi}_0$, $\hat{\gamma}$ are the MLE under $H_0$. 
Then an EM test statistic can be defined as
\begin{align}
EM_n^{(K)} = \max\{M_n^{(K)}(\bm{\beta}_0): \bm{\beta}_0 \in \mathcal{B}^{m_0}\}.
\end{align}
Although the formulation of our EM test statistic shares similarity with that of \citet{Li2010}, the major distinction is that our method treats a regression-type problem that allows a general density form $f$ to relate response $y$ with multiple covariates $\bm{X}$ and $\bm{Z}$, among which the later ones do not determine the mixture structure. In contract, Li and Chen (2010) only handles mean-only models. 
In next section, we will show that under appropriate conditions, the proposed estimators still have desirable asymptotic properties. 

\vspace{-0.5in}
\section{Asymptotic theory}\label{sec:theory}
In this section, we study large-sample properties of the proposed EM test statistics. For simplicity, we define $D = (Y,\bm{X},\bm{Z})$, $D_i = (Y_i, \bm{X}_i,\bm{Z}_i)$. Theorem \ref{tm1} implies that the values of parameters such as $\bm{\alpha}$, $\bm{\beta}$ and $\bm{\theta}$ change at most $o_p(1)$ within each iteration. This result is crucial in obtaining asymptotic distribution of the test statistics under the null hypothesis. The proof of the theorems are given in the Online Supplementary File. 
\begin{theorem}\label{tm1}
Suppose that Conditions (C1)--(C8) listed in the Web Supplementary File hold. Under the null distribution $f(D;\Psi_0,\bm{\gamma}_0)$, the following holds for every $\bm{\beta}_0 \in \mathcal{B}^{m_0}$ and every $k \in \mathbb{N}$,
\begin{align*}
	& \bm{\alpha}^{(k)} - \bm{\alpha}_0 = O_p(n^{-1/2}), ~~ \bm{\beta}^{(k)} - \bm{\beta}_0 = O_p(n^{-1/6}),~~\bm{\theta}_1^{(k)} - \bm{\theta}_0 = O_p(n^{-1/4}),\nn  \\
	&\bm{\theta}_2^{(k)} - \bm{\theta}_0 = O_p(n^{-1/4}), ~~\bm{m}_1^{(k)} = O_p(n^{-1/2}),~~ \bm{\gamma}^{(k)} - \bm{\gamma}_0 = O_p(n^{-1/2}),
	\end{align*}
where $\bm{\theta}_0$, $\bm{\theta}_1^{(k)}$ and $\bm{\theta}_2^{(k)}$ are matrices with $j$-th column as $\bm{\theta}_{0j}$, $\bm{\theta}_{1j}^{(k)}$ and $\bm{\theta}_{2j}^{(k)}$, respectively, and $\bm{m}_1^{(k)} = (\bm{m}_{11}^{(k)},\ldots,\bm{m}_{1m_0}^{(k)})$ with 
\begin{align}
 \bm{m}_{1h}^{(k)} = \beta_h^{(k)} (\bm{\theta}_{1h}^{(k)} - \bm{\theta}_{0h}) +(1- \beta_h^{(k)}) (\bm{\theta}_{2h}^{(k)} - \bm{\theta}_{0h}), ~h=1,\dots,m_0.
\end{align} 
\end{theorem}
We define the following quantities, 
	\begin{align}
	& \Delta_{ih} = \frac{ f(D_i; \bm{\theta}_{0h},\bm{\gamma}_0) -  f(D_i; \bm{\theta}_{0m_0},\bm{\gamma}_0)}{f(D_i; \Psi_0,\bm{\gamma}_0)}, ~~ \bm{Y}_i(\bm{\theta}) = \frac{f'(D_i;\bm{\theta},\bm{\gamma}_0)}{f(D_i; \Psi_0,\bm{\gamma}_0)} \in \mathbb{R}^{p \times 1}~\text{for any}~\bm{\theta} \in \mathbb{R}^{p \times 1}\nn \\
	& \textcolor{blue}{\bm{Z}_i(\bm{\theta}) = \frac{1}{f(D_i)}    \left( \frac{\partial^2 f(D_i)}{\partial \theta_1 \partial \theta_1}, \ldots, \frac{\partial^2 f(D_i)}{\partial \theta_p \partial \theta_p}  \right)^T}, \nn \\
	& \bm{b}_{1i} = (\Delta_{i1},\ldots,\Delta_{i m_0-1}, \bm{Y}_i(\bm{\theta}_{01})^T, \ldots, \bm{Y}_i(\bm{\theta}_{0m_0})^T)^T \in \mathbb{R}^{(m_0-1 + m_0 p) \times 1}, \nn \\
	& \bm{b}_{2i} = (\bm{Z}_i(\bm{\theta}_{01})^{T}, \ldots,\bm{Z}_i(\bm{\theta}_{0m_0})^{T})^T \in \mathbb{R}^{\textcolor{blue}{m_0 p}\times 1},
	\end{align}
	where \textcolor{blue}{$f'(D_i;\bm{\theta},\bm{\gamma}_0)$ is the derivative of $f$ with respect to $\bm{\theta}$, and }sometimes we write $f(D_i; \bm{\theta},\bm{\gamma}_0)$ as $f(D_i)$ for simplicity. \textcolor{blue}{Note that we do not consider the derivatives with respect to $\bm{\gamma}$ in $\bm{b}_{1i}$ and $\bm{b}_{2i}$  because the main idea of the proof is to first evaluate the difference in (penalized) log-likelihood function under $m_0$ subgroups and $ 2 m_0$ subgroups, and then conduct a Taylor expansion for that difference. Since $\bm{\gamma}$ remains the same across different subgroups, it will not have an effect in the expansion, and hence is not included.}
	
	Define covariance matrices $\bm{B}_{jk} = \text{Cov}(\bm{b}_{ji},\bm{b}_{ki})$ for $j,k=1,2$. Let $\tilde{\bm{b}}_{2i} = \bm{b}_{2i} - \bm{B}_{21} \bm{B}_{11}^{-1} \bm{b}_{1i}$, which is orthogonal to $\bm{b}_{1i}$. Apparently, the covariance matrix of $\tilde{\bm{b}}_{2i}$ is $\tilde{\bm{B}}_{22} = \bm{B}_{22} - \bm{B}_{21} \bm{B}_{11}^{-1} \bm{B}_{12} $. Then the next theorem gives an asymptotic expansion result for the EM test statistic. 

\begin{theorem}\label{tm2}
Suppose that Conditions (C1)--(C8) listed in Web Supplementary File hold and $0.5 \in \mathcal{B}$. Under the null distribution $f(D; \Psi_0,\bm{\gamma}_0)$, for any fixed $K$, as $n \rightarrow \infty$, 
\begin{align*}
EM_n^{(K)} = \sup_{\bm{v} \geq 0} \{ 2\bm{v}^T \sum_{i=1}^n \tilde{\bm{b}}_{2i} -n \bm{v}^T \tilde{\bm{B}}_{22}\bm{v} \} +o_p(1),
\end{align*}
where the supremum is taken over the set
\begin{align*}
\{\bm{v} \geq 0\} = \{\bm{v}=(v_1,\ldots,v_{\textcolor{blue}{m_0 p}})^T: v_j \geq 0,j=1,\ldots,\textcolor{blue}{m_0 p} \}.
\end{align*}
\end{theorem}
Based on Theorem \ref{tm2}, we obtain closed-form asymptotic distribution for $EM_n^{(K)}$ as follows. 
\begin{theorem}\label{tm3}
Suppose that Conditions (C1)--(C8) in the Web Supplementary File hold and $0.5 \in \mathcal{B}$. Let $\bm{w}$ be an $\textcolor{blue}{m_0 p}$-dimensional multivariate normal vector with mean zero and covariance $\tilde{\bm{B}}_{22}$. Under the null distribution $f(D; \Psi_0,\bm{\gamma}_0)$, for any fixed $K$, as $n \rightarrow \infty$, 
\begin{align*}
EM_n^{(K)} \rightarrow \sum_{s=0}^{\textcolor{blue}{m_0 p}} a_s \chi_s^2
\end{align*}
in distribution, where $a_s = \text{P}\{\sum_{k=1}^{\textcolor{blue}{m_0 p}}I(\hat{\bm{v}}_k >0) = s\}$, $\hat{\bm{v}}_k$ is the $k$-th element of $\hat{\bm{v}}  = \text{argmax}_{\bm{v} \geq 0} (2\bm{v}^T \bm{w} - \bm{v}^T \tilde{\bm{B}}_{22} \bm{v})$.
\end{theorem}
Theorem \ref{tm3} states that, under the null hypothesis for sufficiently large $n$, $EM_n^{(K)}$ follows a mixture of chi-square distributions. This result agrees with Theorem 3 in \citet{Li2010} with $p=1$. In general, it is difficult to obtain analytic form of weights $a_s$ for $p>1$. \textcolor{blue}{In practice, we can use Monte-Carlo simulations to estimate them following the same procedure as discussed in \citet{Li2010}, that is, we generate $\bm{w}$ from a multivariate normal distribution as described in Theorem \ref{tm3}, then obtain $\hat{\bm{v}}$ by solving the maximization problem using quadratic programming tools (e.g., {\it quadprog} package in R) based on the simulated $\bm{w}$, and obtain the estimated value for $a_s$. This procedure is repeated for $N$ times (in this paper we choose $N = 10000$) and the average value for $a_s$ is used as its final estimate. }

\textcolor{blue}{It is worthy mentioning that the assumption (C8) rules out some commonly used mixture models such as normal location-scale family; see the discussion in Section 2 of the Web Supplementary File for more details. Therefore the asymptotic results in  \citep{chen2020homogeneity,  niu2011testing} do not overlap with ours, e.g.,  \citet{chen2020homogeneity} considers a normal location-scale mixture model, which cannot be covered in our case. It will be of interest to extend the current EM-test framework for those models under general values of $m_0$ in a future work. }

\section{Simulation}\label{sec:simulation}
\subsection{Normal mixture and logistic mixture models}
We conduct simulations to study the finite-sample performance of the proposed testing procedure in terms of both the type-I error rate and power. We consider six data generating scenarios based on normal and logistic mixtures with the number of subgroups $m_0 \in \{1,2,3\}$. We consider the following three simulation scenarios for normal mixture models. 

{\it Scenario 1.} We let the true number of subgroups $m_0 = 1$ and $\alpha = 1$, i.e., there is only one subgroup. The response variable $Y$ is generated from a normal distribution $N(3 X_1 + 5 X_2 + Z,1)$ with covariates $X_1,X_2$ and $Z$ sampled from an uniform distribution on $(0,1)$. 

{\it Scenario 2.} The response variable $Y$ is generated from a mixture of two normal distributions, with two subgroup-specific covariates $X_1,X_2$ and a common-effect covariate $Z$ sampled from an uniform distribution on $(0,1)$. Specifically, $Y$ is generated from the mixture of $.4 N(X_1 + 6X_2 + Z, 1) + .6 N(2X_1 - 6X_2 + Z, 1)$. Here $\bm{\alpha} = (.4, .6)$, the subgroup-specific regression coefficients $\bm{\theta_1} = (1, 6)$ and $\bm{\theta_2} = (2, -6)$, and the common-effect of $Z$ gives $\gamma  = 1$.

{\it Scenario 3.} We consider $m_0 = 3$ with weights $\bm{\alpha} = (.4,.3.,3)^T$. The response variable $Y$ is generated from a mixture of three normal distributions, with $X_1,X_2,Z$ sampled from an uniform distribution on $(0,1)$. We set the subgroup-specific regression coefficients as $\bm{\theta_1} = (1, 6)$, $\bm{\theta_2} = (2, -6)$, $\bm{\theta_3} = (-2,3)$, and the coefficient for $Z$ as $\gamma  = 1$. 
\textcolor{blue}{For all three scenarios, we assume that the error variance is known; so the only unknown parameters are the regression coefficients.}
We let the sample size $n \in \{500,1000,1500\}$ and examine the nominal levels $\alpha$ of $0.01$ and $0.05$, respectively. We repeat the EM test procedure for three iterations ($K=3$) for each case. \textcolor{blue}{To decide the tuning parameter $C$ in the penalty function $p(\cdot)$, we generate a separate dataset under each simulation scenario, consider a range of different values of $C$ and choose the best one that provides the type I errors that are closest to the nominal values. Based on the results summarized in Tables S1-S3 of the Web Supplementary File, we choose $C = 3$ for $m=1$, $C= .8$ for $m=2$, and $C = 2.0$. From those tables we also find that the type I errors are quite robust to different values of $C$. }

We summarize the type-I error based on 5000 Monte-Carlo replications in the first part of  Table~\ref{sim:table1}. It can be seen that type-I error rates are generally close to the nominal levels across different iterations, regardless of different sample sizes and nominal levels. As expected, the type-I errors get closer to the corresponding nominal levels as the sample size goes larger. 


\begin{table} 
	\caption{Type-I errors and power for various sample sizes and nominal levels based on 5000 simulations for mixture-of-normal example.}
	\begin{center}
		\begin{tabular*}{1\textwidth}{@{\extracolsep{\fill}}lcccccccc}
			\hline
 		& \multicolumn{6}{c}{Type I error}  \\ \hline
			&    \multicolumn{2}{c}{$m_0=1$}   &\multicolumn{2}{c}{$m_0=2$}& \multicolumn{2}{c}{$m_0=3$}\\ \cline{2-3} \cline{4-5}   \cline{6-7}  
			$n$  & $\alpha =.01$ &  $\alpha =.05$  & $\alpha =.01$  &  $\alpha =.05$& $\alpha =.01$  &  $\alpha =.05$    \\
			$500$ & .009 & .061  & .012 & .057&.010&.044 \\ \hline  
			$1000$ & .011 & .054  & .012 & .065  &.009& .047 \\ \hline
			$1500$  & .011 & .055  & .011 & .058 &.011 &.046 \\  \hline   
						& \multicolumn{6}{c}{Power (nominal level $= .05$)}  \\ \hline
				&    \multicolumn{2}{c}{$m_0=1$}   &\multicolumn{2}{c}{$m_0=2$}& \multicolumn{2}{c}{$m_0=3$}\\ \cline{2-3} \cline{4-5}   \cline{6-7}  
			$n$ & weak & strong &  weak & strong  & weak & strong     \\
			$500$ & .928 & .999 & .407  & 1.000 & .277 & 1.000 \\ \hline  
			$1000$ & .980 & .995 & .638 & .998 & .474 & 1.000 \\ \hline
			$1500$ & .990 & .999 & .797 & 1.000 & .572 & 1.000 \\ \hline
		\end{tabular*}
		\label{sim:table1}
	\end{center}
\end{table}

To evaluate the power, for each simulation scenario, we consider two cases with a ``weak'' and a ``strong'' level  of heterogeneity in parameter values between subgroups.\\ 
(1) For Scenario 1, we consider two alternatives, a ``weak'' case, with weights $(.8,.2)$ and coefficients $\bm{\theta_1} = (3, 5), \bm{\theta_2} = (3, 3)$; and a ``strong'' case, with weights $(.6,.4)$ and coefficients $\bm{\theta_1} = (3, 5), \bm{\theta_2} = (3, -5)$.\\
(2) For Scenario 2, we consider two alternatives, a ``weak'' case, with weights $(.4, .4, .2)$ and coefficients $\bm{\theta_1} = (1, 6), \bm{\theta_2} = (2, -6), \bm{\theta_3} = (3, -5.5)$; and a ``strong'' case, with weights $(.25,.25,.25,.25)$, and  coefficients $\bm{\theta_1} = (1, 6), \bm{\theta_2} = (2, -6), \bm{\theta_3} = (3, 5),  \bm{\theta_4} = (6, -6)$.  \\
(3) For Scenario 3, we consider two alternatives, a ``weak'' case, with weights $(.2, .2, .3,.3)$ and coefficients $\bm{\theta_1} = (1, 6), \bm{\theta_2} = (1, -4), \bm{\theta_3} = (2, -6), \bm{\theta_4} = (-2,3)$; and a ``strong'' case, with weights $(.2,.2,.1,.2,.2,.1)$, and  coefficients $\bm{\theta_1} = (1, 6), \bm{\theta_2} = (1, -4), \bm{\theta_3} = (2, -6),  \bm{\theta_4} = (-2, 3), \bm{\theta_5} = (-5, -4),  \bm{\theta_6} = (5, 5)$.

The remaining data generation, including response and covariate distributions, is the same as the setting of type-I error examination. We summarize the proportion of rejecting the null hypothesis based on $5000$ simulations in the second part of Table~\ref{sim:table1}. It can be seen that the power of the test increases up to $1$ rapidly, as the data gets more heterogeneous from ``weak'' to ``strong''. It is also intuitive that the power increases with sample size. To save the space, we present the data generation details and the simulation results for logistic mixture model in Section 1.2 of the Web Supplementary File. 
 
 \color{blue}
\subsection{Comparison with other approaches}
We compare the performance of our EM test with that of the structured logistic-normal mixture model proposed in \citet{Shen2015}. The data is simulated in the same way as model (1) in \citet{Shen2015}, i.e., 
\begin{align}
&Y  = \bm{Z }^T(\bm{\beta_1} + \bm{\beta_2} \delta_i)  + \epsilon ,\label{eq1} \\
& P(\delta  = 1) = \exp(\bm{X }^T \gamma)/ \{ 1 + \exp(\bm{X }^T \bm{\gamma})\},\label{eq2} 
\end{align}
where covariates $\bm{X} = (1,x)$, $\bm{Z} = (1,z,x)$; and $x$ is generated from $N(-1,1)$ and $z$ from a Bernoulli$(.5)$ independently. The hypothesis is $m_0 = 1$ versus $m_0 = 2$, i.e., whether there exists a subgroup or not. Following \citet{Shen2015}, we set $\bm{\beta_1} = (1,0,2)^T$ and $\bm{\beta_2} = (0,0,0)^T$ and $\epsilon \sim N(0,0.5^2)$ under the null hypothesis to evaluate the type I error. To evaluate the power, we set $\bm{\beta_2} = (1,a,b)^T$
and $\bm{\gamma} = (1,c)^T$ where $a,b,c$ take values in $\{0.5,1.0\}$ while $\bm{\beta_1}$ and the error distribution remains the same with before. We let the sample size $n \in \{60,100\}$ and summarize the type I error based on 5000 replications and the power based on 1000 replications in Table S8 and S9 of the Web Supplementary File, respectively. For \citet{Shen2015}'s method, we use their results in Tables 2 and 3 of that paper. For both methods, we only compare the results for $K=1$, i.e., after the first EM iteration, because it already achieves a stable estimate. 

Before presenting the result, we would like to emphasize that the simulation scenarios are considered to the advantage of \citet{Shen2015}'s method since the data are exactly generated from their models in \eqref{eq1} and \eqref{eq2}. On the other hand, our method only requires specifying the outcome model in \eqref{eq1}, without the need to impose the logistic model between the subgroup membership $\delta$ and covariates $\bm{X}$ as in \eqref{eq2}. 

From Table S8 and S9, we find that both methods produce type I errors that are very close to the nominal levels, especially when the sample size is large ($n=100$). In terms of power, our method has a comparable performance with that of \citet{Shen2015}, i.e., out of six combinations of $a,b,c$'s values, our method has a power higher than that of \citet{Shen2015} for three times. These findings confirm that our method is capable of correctly identifying the latent subgroup structure under the model mis-specification of the subgroup membership $\delta$. Also, it is worthy mentioning that our method can handle general values of $m_0$ under the null hypothesis while \citet{Shen2015} is only capable of addressing $m_0 = 1$. 

Next we compare our method with the classification tree approach. We generate data from a mixture of two logistic regression models with two subgroup-specific covariates $X_1,X_2$ sampled from an uniform distribution on $(5,10)$, and subgroup-specific regression coefficients $\bm{\theta_1} = (-.1, -.2)$ and $\bm{\theta_2} = (.2, .1)$. We set the sample size $n=500$ and consider three scenarios: (1) the two subgroups are determined by whether $X_2 > 7$ or not;  (2) the two subgroups are determined by whether $X_1 + X_2 > 14$ or not; (3) the two subgroups are randomly distributed with mixture probabilities $\bm{\alpha} = (.6,.4)$. The classification tree is implemented using R package {\it rpart} and pruned based on minimizing the predicted residual error sum of squares under cross validation (``xerror'' option in R function {\it cptable}). We also implement our method to test $H_0: m_0=2$ versus $H_a: m_0=4$ and examine the type-I error rate, since the data can be treated as being generated under $H_0$ with the presence of two subgroups. The result is summarized in Table S10 of the Web Supplementary File based on 1000 replications. It is evident that our method properly preserves the type-I error rates in all the three scenarios. Note that scenarios 1 and 2 can represent the model mis-specification cases (deviating from our data generating model), which advocates the robustness of our method. For classification tree, we present its tree output for one simulated data in each of the three scenarios in Figure S1 of the Web Supplementary File. In those figures, with each child node treated as a subgroup, the tree method tends to overestimate the number of subgroups when the subgroup classification rule is complex (Scenario 2 and 3). This observation is not surprising because trees are designed mainly for prediction, rather than learning the subgroup structure.

\color{black}

\section{Real data analysis}
A multi-center validation study sponsored by the Early Detection Research Network (EDRN) has been conducted for early detection of prostate cancer. This large-scale study aims to evaluate clinical utility of a urine-based biomarker, prostate cancer antigen 3 (PCA3), for assessment of prostate cancer risk. From December 2009 to June 2011, this validation trial enrolled 928 men who received a prostate biopsy for some standard indications. Among all, 859 men are evaluable with available baseline demographic and clinical characteristics, 47.8\% of whom were diagnosed with prostate cancer. In addition, measurements of several known important biomarkers for prostate cancer, such as Prostate-Specific Antigen (PSA), PCA3, T2-ERG gene fusion, and a composite marker--Prostate Cancer Prevention Trial (PCPT) prostate cancer risk calculator, are also available on these subjects. Prostate cancer heterogeneity is well known, e.g., T2-ERG gene fusion is found in about 50\% of prostate cancer patients. Our interest is to incorporate patient heterogeneity in classifier construction based on biomarkers, for the goal of improving early diagnosis of prostate cancer.   
 \textcolor{black}{Due to data sharing confidentiality, in our analysis, we first implemented a small perturbation of the original data, and then performed all the analysis based on the perturbed data set. Demonstration of applicability of our method can still be achieved with this practice.}

\begin{figure}[h]
	\begin{tabular}{cc}
	\includegraphics[width = 0.5\textwidth]{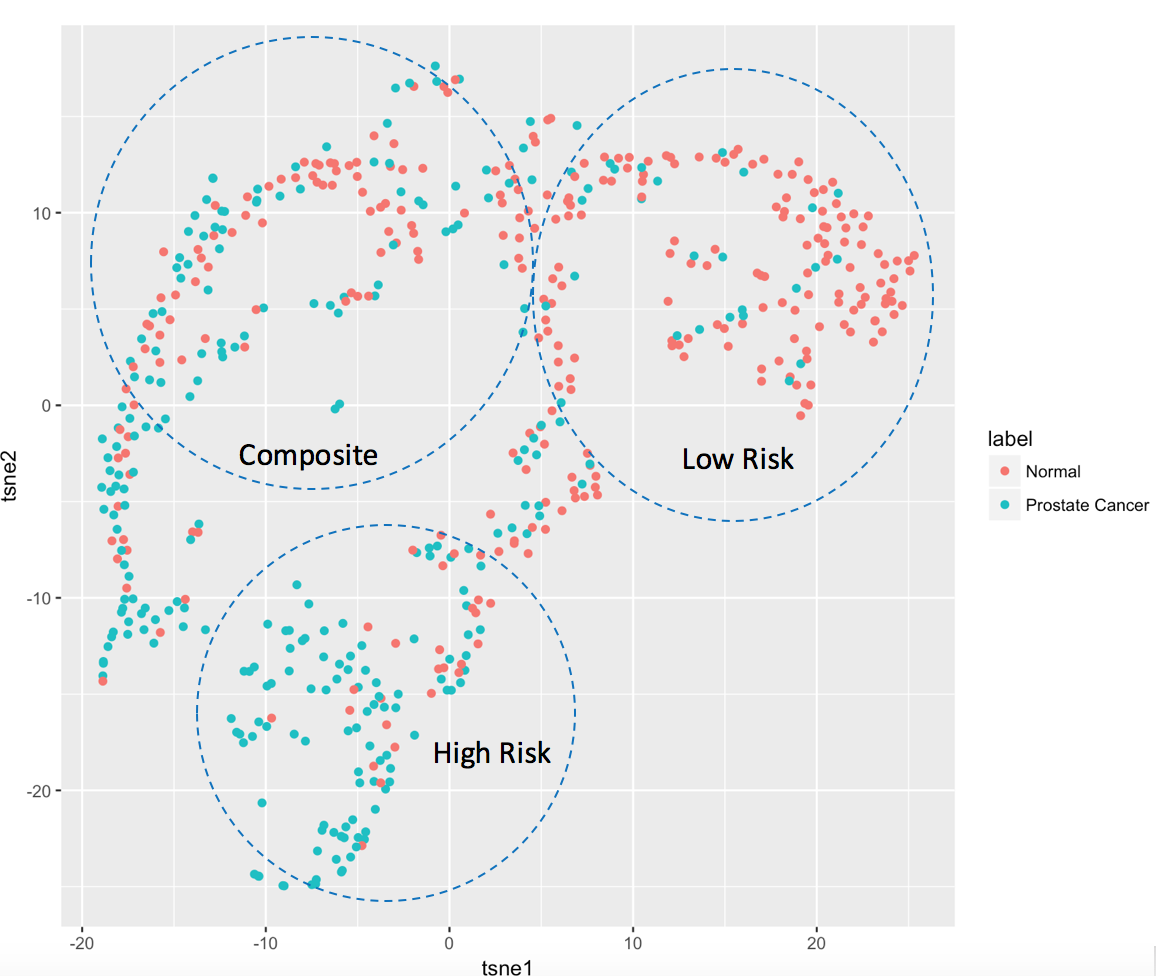}  & 	\includegraphics[width = 0.5\textwidth]{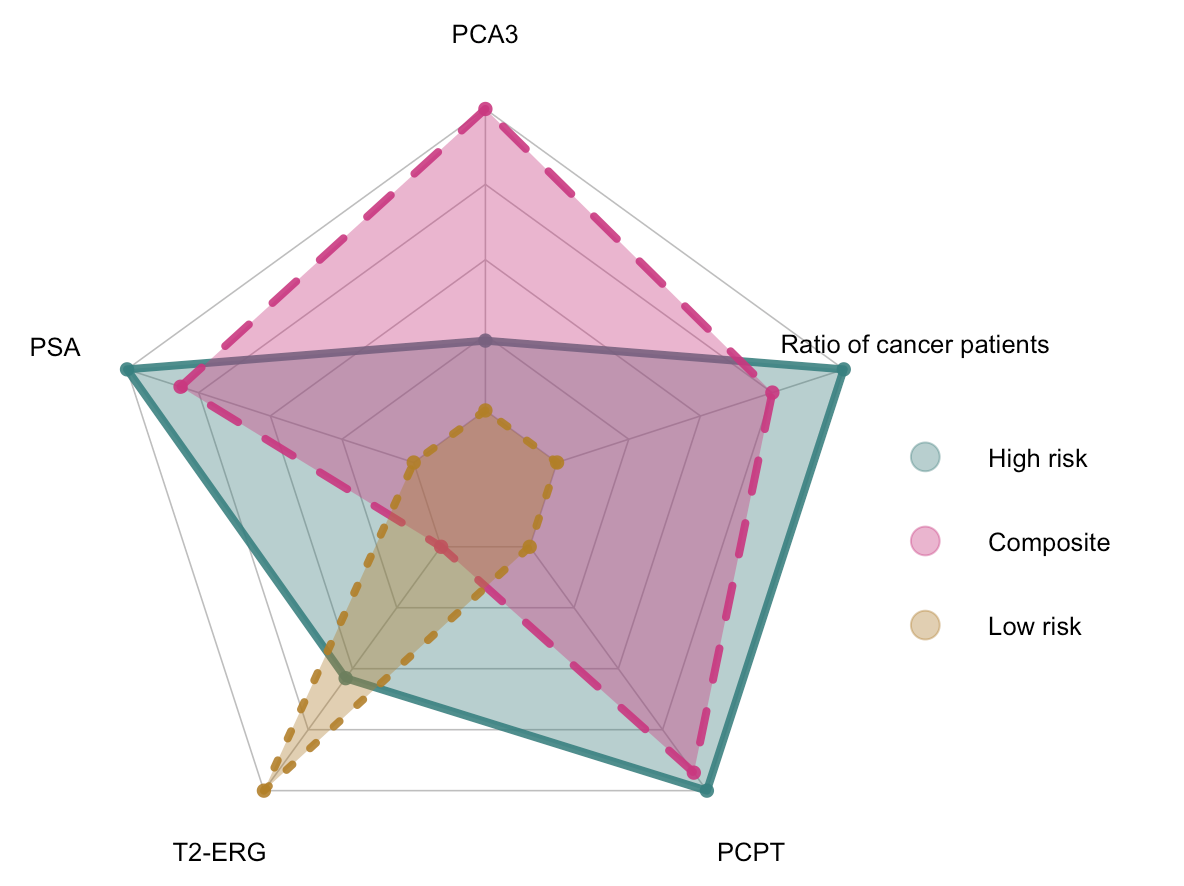}  \\ \\
	\end{tabular}  
	\hskip 2in
		\caption{(a) t-SNE visualization of covariates over three groups; and (b) Radar plot of the four biomarkers and cancer occurrence rates over three patient subgroups}
	\label{fig:real_tsne}
\end{figure}

First, we would like to see if empirically heterogeneity is observable in the association between biomarkers and prostate cancer status. \textcolor{blue}{We consider a t-distributed Stochastic Neighbor Embedding (t-SNE), which is a widely used visualization and dimension reduction method proposed by \cite{maaten2008visualizing}. Unlike classical principal component analysis (PCA), t-SNE is capable of embedding the high dimensional data points into a low dimensional space via a nonlinear transformation, and works quite well for many biological and medical studies.} Figure~\ref{fig:real_tsne} (a) is a scatter plot originated from all four biomarkers and cancer diagnosis. In the projected two-dimensional space (derived from the original four biomarkers) where the x-axis and y-axis represent the first two leading directions, we obtain a clear visualization of three groups corresponding to low, high, and medium cancer risks. Such heterogeneity can be easily ignored by commonly used statistical models, including logistic regression. 

\begin{figure}[h]
	\hskip -2cm
	\begin{tabular}{c}
		\includegraphics[width = 1\textwidth]{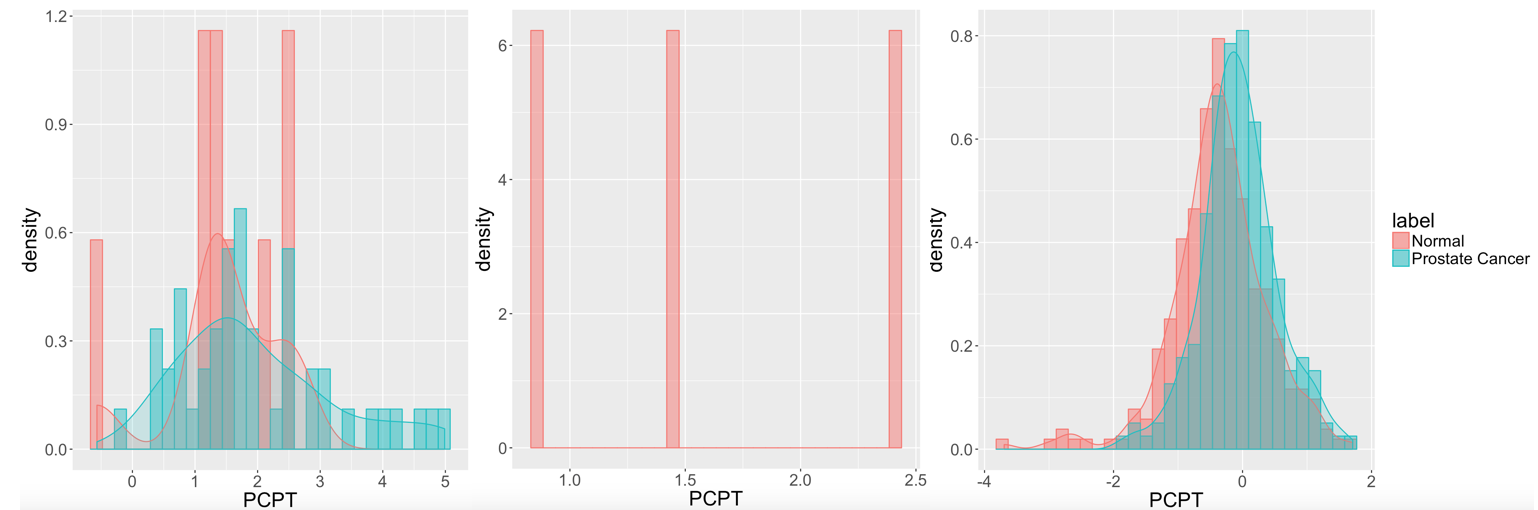}\\
		\includegraphics[width = 1\textwidth]{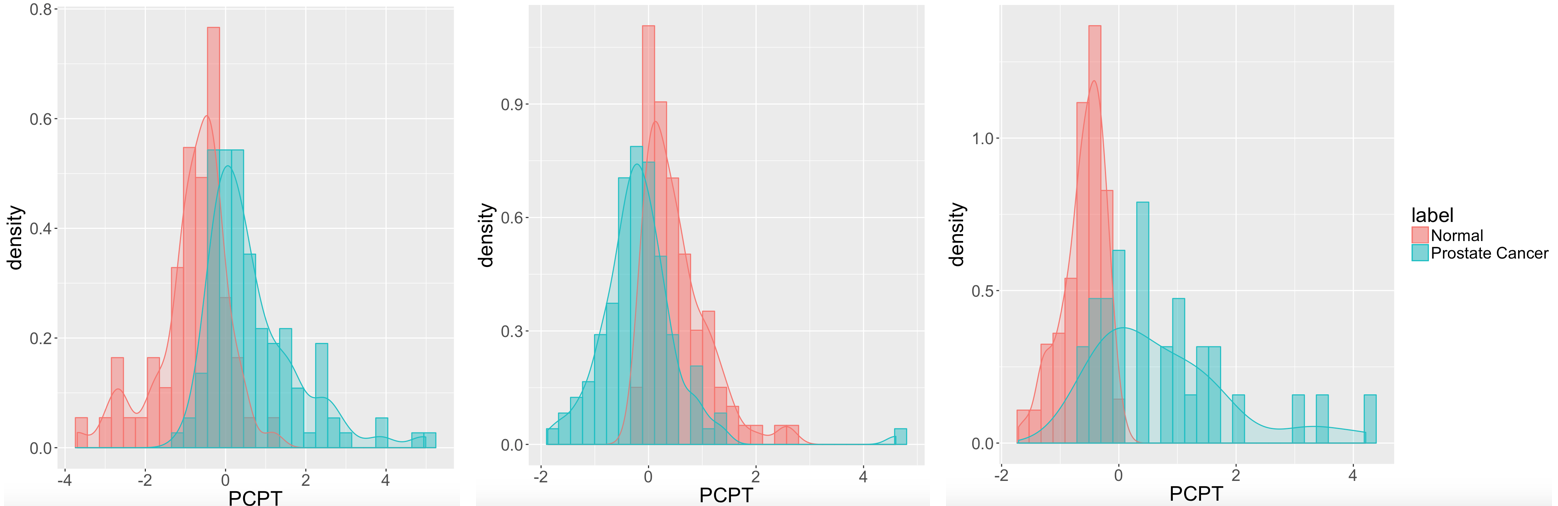} \\ \\
	\end{tabular} \hskip 4in
	\caption{Distribution plots of PCPT over diagnosis groups (normal / prostate cancer) and subgroups. Those subgroups are obtained from K-means (top) and the proposed approach (bottom).}
	\label{fig:real_hist_emsubgroup}
\end{figure}

From an alternative angle, we conduct K-means clustering using these four biomarkers, which results in three patient subgroups. We use PCPT as an example, and respectively plot its distributions in normal and cancer groups across all the subgroups for comparison in Figure~\ref{fig:real_hist_emsubgroup}. 
Despite the insufficient samples from one subgroup, it is observed (from the top row) that the distributions of PCPT in both normal and cancer groups behave distinctively among subgroups. The multi-mode pattern in the leftmost subgroup suggests an inconsistent association with disease when compared to the one on the rightmost. The similar observation is also made for the other three biomarkers. Moreover, the imbalanced subgroup results from K-means can be easily observed from the middle figure.

These empirical exploration indicates existence of nontrivial heterogeneity in terms of association between biomarkers and prostate cancer risk. Classical statistical models are likely incapable of handling such data complexity, and potentially result in misleading scientific conclusions. To address the heterogeneity issue, we implement the proposed method that was described in the previous section. Specifically, \textcolor{blue}{we consider a mixture of logistic regression models and} perform a sequential hypothesis testing procedure on the number of subgroups, $H_0: m_0=m$ versus $H_a: m_0=2m$, to test the existence of subgroups ($m=1$) and identify the optimal number of subgroups if $m>1$. It starts from $m=1$ and continues to increase by 1 until the null hypothesis at the current value of $m$ cannot be rejected. 
Table~\ref{tab:real_pvalues} shows the p-values of the sequential tests up to $m=3$ , at which the test cannot be rejected. For each test, we show the result based on different numbers of iterations ($K=1,2,3$). The similar results obtained among different values of $K$ indicate that the test is quite robust to the number of iterations and a small number of $K$ is typically sufficient. From the sequential testing procedure, we declare that the data set most likely contains three subgroups in terms of association between biomarkers and cancer occurrence. We notice that cancer occurrence rates of .186, .548, and .668, respectively obtained for the three estimated subgroups, actually coincide with the pattern shown in Figure~\ref{fig:real_tsne} (a).  
We also compare the distributions of the biomarker {\it PCPT} in three subgroups identified by K-means in the upper panels of Figure~\ref{fig:real_hist_emsubgroup} to those identified by the proposed EM-test in the lower panels of the same Figure. 
It is clear that directly associating the response with covariates to identify patient subgroups in the proposed method results in more meaningful disease-related subgroup construction than K-mean clustering of mere covariates. 

Moreover, Figure~\ref{fig:real_tsne} (b) displays a radar plot of the data over the three subgroups. A radar plot is typically used to present multivariate data in a two-dimensional format. Herein,  each vertex of the polygon represents one covariate where the length of edge measures the mean value corresponding to the three subgroups. The colors indicate different subgroups. 
Clearly, the discrepancy of the biomarker expression and cancer occurrence rate over the three subgroups is manifested. It is notable that T2-ERG has the largest value in the low risk group among all the three groups, while PCA3 has the smallest value compared to all the other three biomarkers in the high risk group. These interesting findings can be easily neglected by those commonly used statistical methods. 
\begin{table}[h]
	\caption{P values of the proposed test results (C=.8) in the prostate cancer study.}
	\begin{center}
		\begin{tabular*}{1\textwidth}{@{\extracolsep{\fill}}lcccc}
			\hline
			Test & Iteration 1 &Iteration 2 & Iteration 3  \\\hline
			$1~\text{vs}~2$ & .029 & .029 & .029  \\ \hline  
			$2~\text{vs}~4$ & .023 & .022 & .021  \\ \hline
			$3~\text{vs}~6$& .099 & .099 & .099  \\ \hline
		\end{tabular*}
	\end{center}
	\label{tab:real_pvalues}
\end{table}

Furthermore, we are interested in examining the performance of clinical response prediction (cancer status). We compare our method to several other competitors---logistic regression, decision tree, random forest, random grouping, and K-means. The method of ``grouping \& logistic regression" is to separately fit logistic regression models to identified subgroups by a grouping method. We use the performance assessment metrics of accuracy ($\frac{\text{true positives} + \text{true negatives}}{\text{total population}}$), 
precision ($\frac{\text{true positives}}{\text{true positives} + \text{false positives}}$), recall ($\frac{\text{true positives}}{\text{true positives} + \text{false negatives}}$), F1 score ($\frac{2 * \text{precision} * \text{recall}}{\text{precision} + \text{recall}}$), and AUC of the ROC curve to make comparisons among all the methods. Table~\ref{tab:real_CV} summarizes five-fold cross validation results. It is clear that our approach outperforms all the other methods comprehensively. When comparing between the grouping-based approaches, K-means grouping performs slightly better than random grouping. Note that the naive K-means grouping coupled with logistic regression already provides some advantage over direct logistic regression in terms of precision and recall, which also suggests that the heterogeneity nature of the data should be taken into account. 

\begin{table}[h]
	\caption{Five fold cross validation results comparing between the proposed approach and competitors}
	\begin{center}
		\begin{tabular*}{1\textwidth}{@{\extracolsep{\fill}}lcccccc}
			\hline
			Method & Accuracy & Precision & Recall &  F1 Score & AUC \\\hline
			$\text{Our Approach}$ & .989 & .977 & 1.000& .988& .995 \\ \hline
			$\text{Logistic Regression}$ & .728 & .687 & .883 &.771&.834 \\ \hline  
			
			$\text{Decision Tree}$ & .707 & .720 & .725 &.720 &NA \\ \hline  
				
			$\text{Random Forest}$ & .708 & .723 & .712 &.716&NA \\ \hline  
			
           	$\text{Random Grouping \& Logit Regression}$ & .685 & .666 & .872 &.752 &.795 \\ \hline  
           	$\text{K-means Grouping \& Logit Regression}$ & .711 & .690 & .891 & .777 & .810 \\ \hline  
		\end{tabular*}
	\end{center}
	\label{tab:real_CV}
\end{table}

At last, we show the parameter estimation results of our approach and logistic regression, respectively, in Table~\ref{tab:coefficients}. The parameter estimates of the four biomarkers in the three subgroups identified by our method are separately shown in the three columns to the left, while those obtained by logistic regression are displayed in the rightmost column. An intriguing finding is that the low risk group (group 3) displays similar parameter estimates as those of logistic regression, from which the patterns in groups 1 and 2 are very different. Particularly, high risk and low risk groups are the most different in terms of all the parameter estimates. On the other hand, PSA and PCPT act as the main contributors to separate high risk group from composite group. Failure of logistic regression in capturing the distinctive structure hidden in subgroups 1 and 2 leads to the inferior performance manifested in Table~\ref{tab:real_CV}.  
Again, the result suggests that, without taking into account patient heterogeneity, classical statistical models may underestimate complexity of the data and potentially result in misleading findings. 
\begin{table}[h]
	\caption{Estimates of coefficients from the proposed approach and logistic regression}
	\begin{center}
		\begin{tabular*}{1\textwidth}{@{\extracolsep{\fill}}lccccc}
			\hline
		  Covariate &  \multicolumn{3}{c}{Our approach}  & Logistic   \\  \cline{2-4} 
				 & Group 1 (High risk) & Group 2 (Composite) & Group 3 (Low risk)&  regression  \\ \hline
			PCA3 & -.057 & -.171 & 1.762 & 1.106 \\ \hline  
			PSA & .627 & -.083 & -.199 &-.094 \\ \hline
			T2-ERG & .747 & .573 & 2.492 & 2.582\\ \hline
			PCPT & -.304 & .819 & .994 & .799\\ \hline
		\end{tabular*}
	\end{center}
	\label{tab:coefficients}
\end{table}

\section{Discussion}
In this paper, we propose a EM-based test on existence and the number of subgroups in a generalized linear regression framework with multiple covariates. We also establish the asymptotic distribution result of the test statistic under the null hypothesis. Note that existing methods mainly focus on mean-only-based subgroup tests without accounting for any covariates. This development intends to address the well-recognized issue of disease heterogeneity using biomarker profiles and other important clinical factors in a broad biomedical applications. As demonstrated by the biomarker-based validation study for early detection of prostate cancer in the paper, the proposed method can potentially have meaningful clinical implications, including the promise of improving accuracy of disease diagnosis by accounting for patient non-homogeneity.  

\textcolor{blue}{In the proposed EM test framework, the choice of penalty function $p(\cdot)$ plays an important role since it controls the level of over-fitting under the alternative hypothesis. In our numerical studies, we use the same penalty function with the one in \citet{Li2010} for convenience, although \citet{Li2010} did not consider the covariates. It is of interest to design a new penalty function that incorporates the covariate information in a future work.}


\backmatter


\section*{Acknowledgements}
The authors thank the editor, the associate editor and referees for their constructive comments on the paper. Shen's research is partially supported by the Simons Foundation Award 512620 and NSF DMS 1509023. Hu's research is partially supported by the National Institute of Health (NCI 5P30 CA013696, NIAID 1R01 AI143886, NIH/NCI 1R01 CA219896). 
  \bibliographystyle{biom} 
\bibliography{ref_mixt}
\section*{Supporting Information}
Web Appendices, Tables, and computational code referenced in Sections \ref{sec:theory}  and \ref{sec:simulation} are available with this paper at the Biometrics website on Wiley Online Library.
\end{document}


\title{Supplementary File for ``Addressing patient heterogeneity in disease predictive model development''}
  
 \author{Xu Gao, Weining Shen, Jing Ning, Ziding Feng, and Jianhua Hu
    }
    \date{}
\maketitle

\section{Additional numerical results}
We present additional simulation results from Section 4 of the main paper here. 
\subsection{Additional simulation results for normal mixture models}
We provide the tuning parameter $C$ selection results for normal mixture model with $m_0 = 1,2,3$ in Tables S1--S3. In each table, the selected value for $C$ is highlighted in bold text, e.g., $C = 3.0$ in Table S1; and this value is chosen based on the type I error (closest to the nominal levels).

\begin{table}[H]\label{tab:sim_supp1}\caption{Tuning parameter $C$ selection: rejection proportion for mixture-of-normal example with $m_0=1$.}	
\begin{center}
\begin{tabular}{  c|c c c  } 
 \hline
	C & $\alpha=.01$ & $\alpha=.05$ & $\alpha=.1$ \\\hline
		0.2 & .013 & .078 & .145  \\   
			0.4 & .012 & .071  &  .133 \\  
			0.6 & .010 & .064 &  .127 \\  
	    	0.8 & .010 & .061  & .122   \\  
			1.0 & .010 & .060 & .115   \\  
	     	1.5 & .009 & .058 &  .110 \\  
			2.0 & .008 & .058 & .106  \\  
			\textbf{3.0} & .008 & .056 & .105  \\  
			8.0 & .007 & .054 & .101  \\
	    	12.0 & .007 & .053 & .100  \\
 \hline
\end{tabular}
\end{center}
\end{table}
 
\begin{table}[H]\label{tab:sim_supp2}\caption{Tuning parameter $C$ selection: rejection proportion for mixture-of-normal example with $m_0=2$.}	
\begin{center}
\begin{tabular}{  c|c c c  } 
 \hline
	C & $\alpha=.01$ & $\alpha=.05$ & $\alpha=.1$ \\\hline
		0.2 & .017 & .065 & .137  \\   
			0.4 & .016 & .060  &  .117 \\  
			0.6 & .014 & .055 &  .107 \\  
	    	\textbf{0.8} & .013 & .051  & .099   \\  
			1.0 & .012 & .045 & .088   \\  
	     	1.5 & .009 & .039 &  .081 \\  
			2.0 & .008 & .033 & .072  \\  
		 \hline
\end{tabular}
\end{center}
\end{table}

\begin{table}[H]\label{tab:sim_supp3}\caption{Tuning parameter $C$ selection: rejection proportion for mixture-of-normal example with $m_0=3$.}	
\begin{center}
\begin{tabular}{  c|c c c  } 
 \hline
	C & $\alpha=.01$ & $\alpha=.05$ & $\alpha=.1$ \\\hline
	    	0.2 & .021 & .094 & .175  \\   
			0.4 & .019 & .084  &  .157 \\  
			0.6 & .014 & .082 &  .140 \\  
	    	0.8 & .013 & .079  & .132   \\  
			1.0 & .012 & .072 & .123   \\  
	     	1.5 & .010 & .066 & .106 \\  
			\textbf{2.0} & .008 & .056 & .096  \\  
			3.0 & .007 & .046 & .080  \\  
			5.0 & .006 & .038 & .066  \\ 
 \hline
\end{tabular}
\end{center}
\end{table}

\subsection{Results for logistic mixture models}
Here we present the simulation results for mixture of logistic regression models with $m_0 = 1,2,3$ here. We consider three simulation scenarios. 

{\it Scenario 1.} We let the true number of subgroups $m_0 = 1$ and $\alpha = 1$, i.e., there is only one subgroup. The response variable $Y$ is generated from a logistic regression model with two covariates $X_1,X_2$ sampled from an uniform distribution on $(5,10)$, and the coefficient is $\bm{\theta} = (.4,.6)$.

{\it Scenario 2.} The response variable $Y$ is generated from a mixture of two logistic regression models, with two subgroup-specific covariates $X_1,X_2$ sampled from an uniform distribution on $(5,10)$. Here $\bm{\alpha} = (.6, .4)$, the subgroup-specific regression coefficients are $\bm{\theta_1} = (-.1, -.2)$ and $\bm{\theta_2} = (.2, .1)$.

{\it Scenario 3.} We consider $m_0 = 3$ with weights $\bm{\alpha} = (.4,.3.,3)^T$. The response variable $Y$ is generated from a mixture of three logistic regression models, with $X_1,X_2$ sampled from an uniform distribution on $(5,10)$. We set the subgroup-specific regression coefficients as $\bm{\theta_1} = (-.1, -.2)$, $\bm{\theta_2} = (0, .3)$, $\bm{\theta_3} = (.3,0)$.

We let the sample size $n$ take the values of $500$, $1000$, and $1500$ and examine the nominal levels $\alpha$ of $0.01$ and $0.05$, respectively. We repeat the EM test procedure for three iterations ($K=3$) for each case. To decide the tuning parameter $C$ in the penalty function $p(\cdot)$, we generate a separate dataset under each simulation scenario, consider a range of different values of $C$ and choose the best one that provides the type I errors that are closest to the nominal values. Based on the results summarized in Tables S4-S6, we choose $C = 1.8$ for $m=1$, $C= 1.0$ for $m=2$, and $C = 2.0$ for $m=3$. From those tables we also find that the type I errors are quite robust to different values of $C$. 

We summarize the type-I error based on 5000 Monte-Carlo replications in the first part of  Table~S7. It can be seen that type-I error rates are generally close to the nominal levels across different scenarios, regardless of different sample sizes and nominal levels.

\begin{table}[H]\label{tab:sim_supp4}\caption{Tuning parameter $C$ selection: rejection proportion for mixture-of-logistic example with $m_0=1$.}	
\begin{center}
\begin{tabular}{  c|c c c  } 
 \hline
	C & $\alpha=.01$ & $\alpha=.05$ & $\alpha=.1$ \\\hline
		0.4 & .019 & .079 & .151  \\   
			0.8 & .014 & .066  &  .122 \\  
			1.4 & .013 & .055 &  .106 \\  
	    	\textbf{1.8} & .013 & .050  & .101  \\  
			2.2 & .013 & .048 & .098   \\  
			3.0 & .012 & .045 & .089  \\  
		 \hline
\end{tabular}
\end{center}
\end{table}

\begin{table}[H]\label{tab:sim_supp5}\caption{Tuning parameter $C$ selection: rejection proportion for mixture-of-logistic example with $m_0=2$.}	
\begin{center}
\begin{tabular}{  c|c c c  } 
 \hline
	C & $\alpha=.01$ & $\alpha=.05$ & $\alpha=.1$ \\\hline
		0.1 & .013 & .067 & .131  \\   
			0.5 & .013 & .062  &  .112 \\  
			.9 & .012 & .051 &  .096 \\  
	    	\textbf{1.0} & .012 & .049  & .095  \\  
			1.5 & .010 & .045 & .083   \\  
			2.0 & .010 & .036 & .078  \\  
		 \hline
\end{tabular}
\end{center}
\end{table}

\begin{table}[H]\label{tab:sim_supp6}\caption{Tuning parameter $C$ selection: rejection proportion for mixture-of-logistic example with $m_0=3$.}	
\begin{center}
\begin{tabular}{  c|c c c  } 
 \hline
	C & $\alpha=.01$ & $\alpha=.05$ & $\alpha=.1$ \\\hline
		0.2 & .015 & .097 & .202  \\   
			0.6 & .013 & .081  &  .166 \\  
			1.0 & .008 & .054 &  .120 \\  
	     	1.5 & .006 & .045 & .108   \\  
			1.9 & .006 & .043 & .101 \\ \textbf{2.0} & .005 & .042  & .099  \\   
		 \hline
\end{tabular}
\end{center}
\end{table}

\begin{table}[H]
	\caption{Type-I errors and power for various sample sizes and nominal levels based on 5000 simulations for mixture-of-logistic example.}
	\begin{center}
		\begin{tabular*}{1\textwidth}{@{\extracolsep{\fill}}lcccccccc}
			\hline
 		& \multicolumn{6}{c}{Type I error}  \\ \hline
			&    \multicolumn{2}{c}{$m_0=1$}   &\multicolumn{2}{c}{$m_0=2$}& \multicolumn{2}{c}{$m_0=3$}\\ \cline{2-3} \cline{4-5}   \cline{6-7}  
			$n$  & $\alpha =.01$ &  $\alpha =.05$  & $\alpha =.01$  &  $\alpha =.05$& $\alpha =.01$  &  $\alpha =.05$    \\
			$500$ & .007 & .033  & .012 & .049 &.003 &.034 \\ \hline  
			$1000$ & .009 & .0375  & .012 & .067  &.009& .048 \\ \hline
			$1500$  & .013 & .050  & .013 & .075 &.010 &.062 \\  \hline   
						& \multicolumn{6}{c}{Power (nominal level $= .1$)}  \\ \hline
				&    \multicolumn{2}{c}{$m_0=1$}   &\multicolumn{2}{c}{$m_0=2$}& \multicolumn{2}{c}{$m_0=3$}\\ \cline{2-3} \cline{4-5}   \cline{6-7}  
			$n$ & weak & strong &  weak & strong  & weak & strong     \\
			$500$  & .442 & .472 &.172 & .367 &.121 &.131 \\ \hline  
			$1000$ & .608 & .686 &.273 & .496 &.224 &.251 \\ \hline
			$1500$ & .681 & .834 &.455 & .620 &.277 &.354 \\ \hline
		\end{tabular*}
		\label{table7}
	\end{center}
\end{table}

To evaluate the power, for each simulation scenario, we consider two cases that represent different levels of heterogeneity in terms of the parameter value differences between the subgroups. 
\begin{itemize}
\item For Scenario 1, we consider two alternatives, a ``weak'' case, with weights $(.5,.5)$ and coefficients $\bm{\theta_1} = (-.4, -.2), \bm{\theta_2} = (0, 0)$; and a ``strong'' case, with weights $(.9,.1)$ and coefficients $\bm{\theta_1} = (-.4, -.2), \bm{\theta_2} = (.2, .4)$.
\item For Scenario 2, we consider two alternatives, a ``weak'' case, with weights $(.1, .1, .1,.7)$ and coefficients $\bm{\theta_1} = (-.2, -.4), \bm{\theta_2} = (-.2, 0), \bm{\theta_3} = (-.2, .2), \bm{\theta_4} = (.4, .2)$; and a ``strong'' case, with weights $(.25,.25,.25,.25)$, and  coefficients $\bm{\theta_1} = (-.1,-.2), \bm{\theta_2} = (-.1,0), \bm{\theta_3} = (1,0),  \bm{\theta_4} = (.2,.2)$.  
\item For Scenario 3, we consider two alternatives, a ``weak'' case, with weights $(.2,.2,.2,.2,.1,.1)$ and coefficients $\bm{\theta_1} = (-.1, -.2), \bm{\theta_2} = (0,.3), \bm{\theta_3} = (.3,0), \bm{\theta_4} = (.1,-.2),\bm{\theta_5} = (-.2,.2), \bm{\theta_6} = (-.3,0)$; and a ``strong'' case, with weights $(.1,.2,.2,.2,.2,.1)$, and  coefficients $\bm{\theta_1} = (-.1, -.2), \bm{\theta_2} = (.1,-.2), \bm{\theta_3} = (-.3,.1), \bm{\theta_4} = (-.1,.1),\bm{\theta_5} = (0,.3), \bm{\theta_6} = (.2,.3)$.  
\end{itemize}
The remaining data generation, including response and covariate distributions, is the same as the setting of type-I error examination. We summarize the proportion of rejecting the null hypothesis based on $1000$ simulations in the second part of Table~S7. It can be seen that the power of the test increases up rapidly for $m_0 = 1$ and $2$, but does not perform as well for $m_0 = 3$. There are several possible explanations. In general, the mixture of logistic regression model is more difficult to fit compared to mixture models with a continuous outcome, and hence requires a much larger sample size to achieve a comparable performance with the normal mixture model. When evaluating the power for $m_0 = 3$, the data is simulated from six subgroups, which makes the sample size for each subgroup very small. Therefore it becomes more challenging to detect the difference across different subgroups. Another possible explanation is due to the violation of Assumption (C8) as discussed in Section 2 of the Web Supplementary File. 
 
\subsection{Comparison with \citet{Shen2015}'s method}
We compare the performance of our EM test with the method proposed in \citet{Shen2015} in Table S8 and S9 here.

\begin{table}[H]
	\caption{Type-I errors for our method and \citet{Shen2015}'s method.}
	\begin{center}
		\begin{tabular*}{1\textwidth}{@{\extracolsep{\fill}}lcccccccc}
			\hline
 			&    \multicolumn{3}{c}{Our method}   &\multicolumn{3}{c}{\citet{Shen2015}} \\ \cline{2-4} \cline{5-7}    
			$n$  & $\alpha =.01$ &  $\alpha =.05$  & $\alpha =.1$  &  $\alpha =.01$& $\alpha =.05$  &  $\alpha =.1$    \\
			$60$ & .011 & .040  & .076 & .012 &.045 &.094 \\ \hline  
			$100$ & .010  & .045  & .088 & .011  &.050& .100 \\ \hline
		\end{tabular*}
		\label{table8}
	\end{center}
\end{table}

\begin{table}[H]\label{table8}\caption{Power for for our method and \citet{Shen2015}'s method with nominal level $=.05$.}	
\begin{center}
\begin{tabular}{  c|c c c |c c } 
 \hline
	$n$ & $a$ & $b$ & $c$ & Our method & \citet{Shen2015} \\\hline
		60 &  0.5 & 1 & 1 & .346 & .730 \\   
		  &  0.5 & 0 & 1 & .386 & .186 \\  
		  &  0.5 & 1 & 0 & .336 & .302 \\   
		   &  1.0 & 1 & 1  & .490 & .820 \\   
		    &  1.0 & 0 & 1 & .691 & .504  \\   
		     &  1.0 & 1 & 0 & .448 & .454 \\   
	   100 &  0.5 & 1 & 1 & .525  & .970 \\   
		  &  0.5 & 0 & 1 & .584 & .376  \\  
		  &  0.5 & 1 & 0 & .506 & .420 \\   
		   &  1.0 & 1 & 1  & .704 & .980  \\   
		    &  1.0 & 0 & 1 & .887 & .700 \\   
		     &  1.0 & 1 & 0 &.659 & .654  \\  
		 \hline
\end{tabular}
\end{center}
\end{table}

\subsection{Comparison with tree method}

\begin{figure}[h]
	\begin{tabular}{ccc}
	\includegraphics[width = 0.33\textwidth]{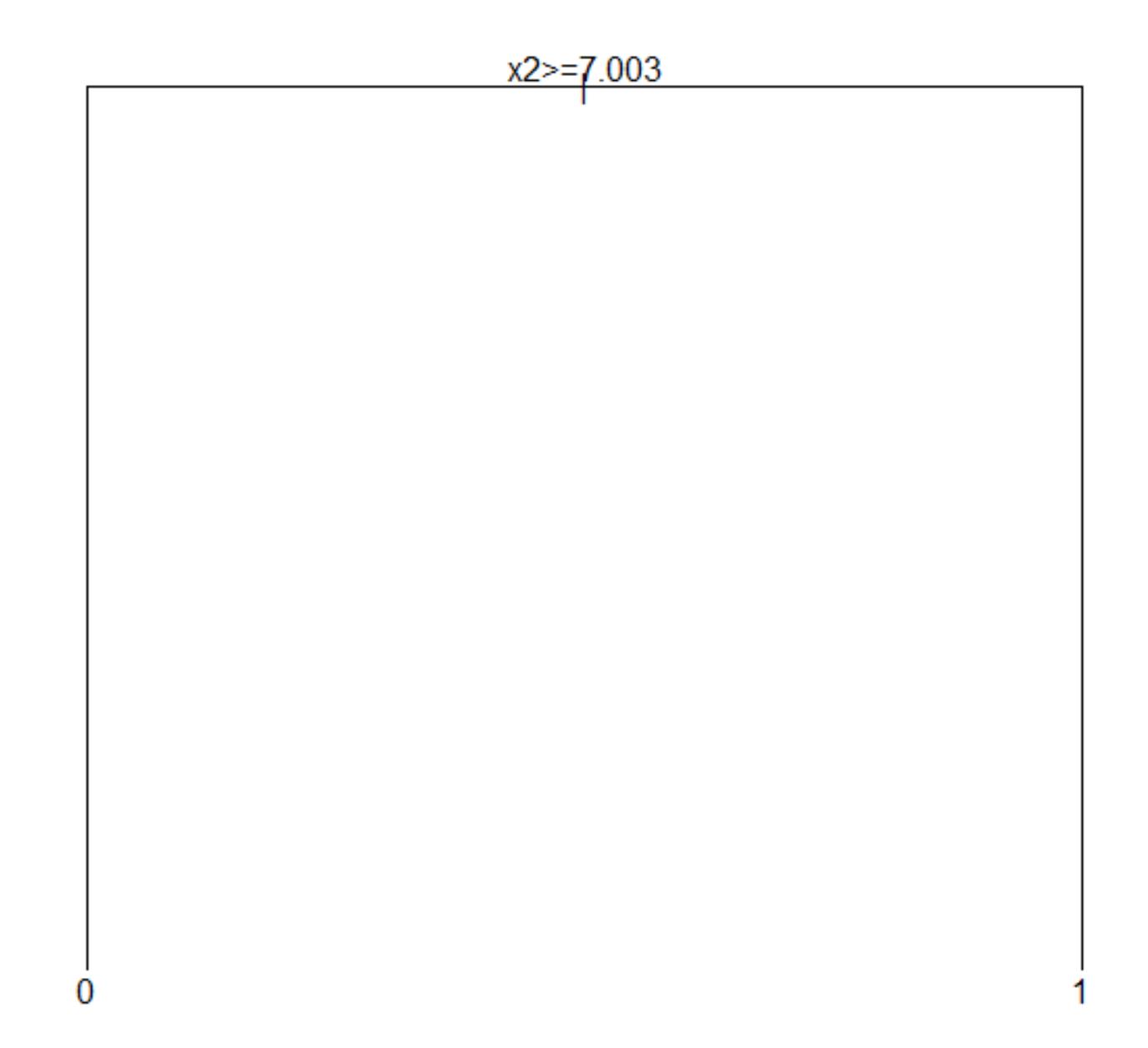}  & 	\includegraphics[width = 0.33\textwidth]{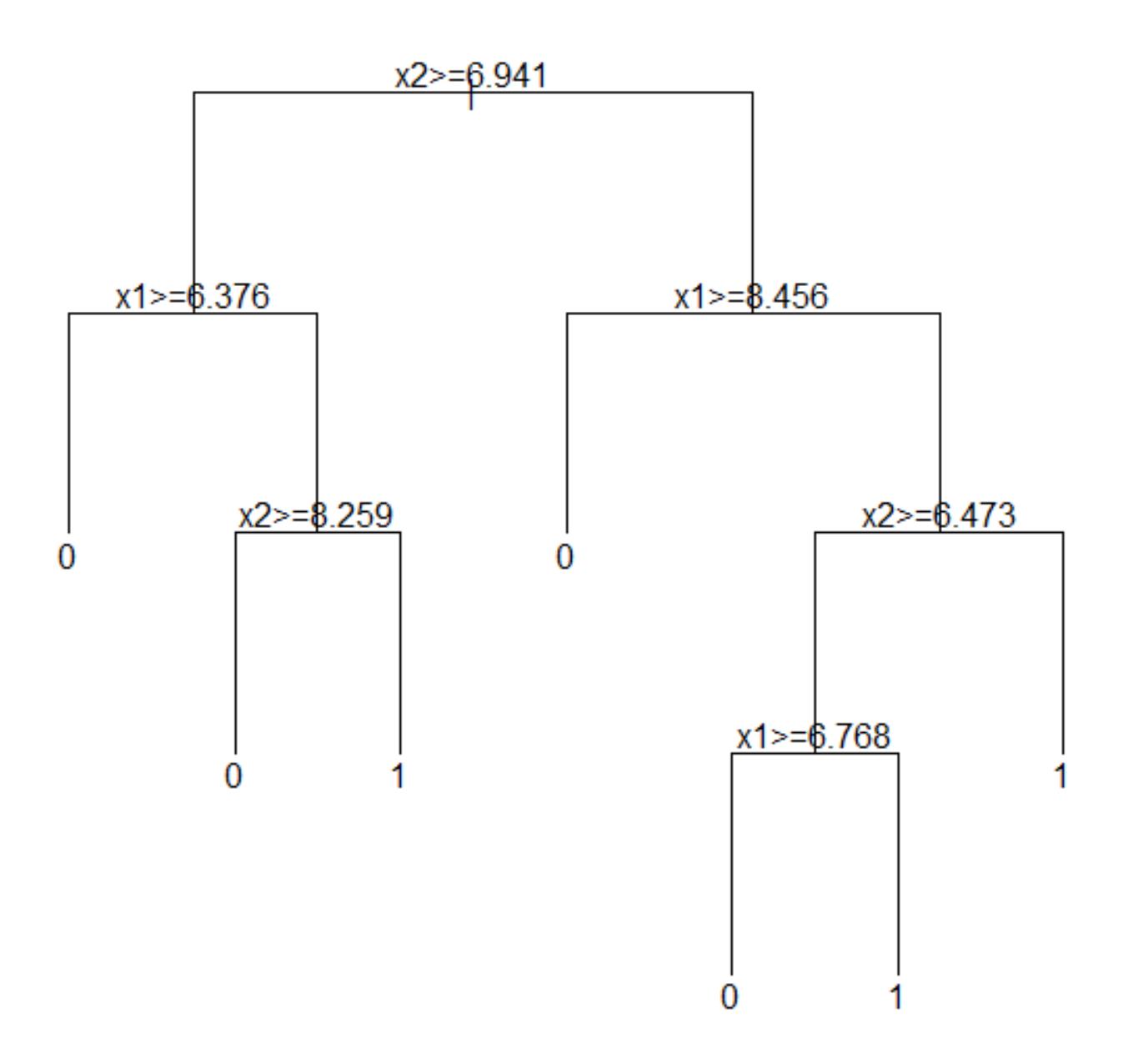} & 	\includegraphics[width = 0.33\textwidth]{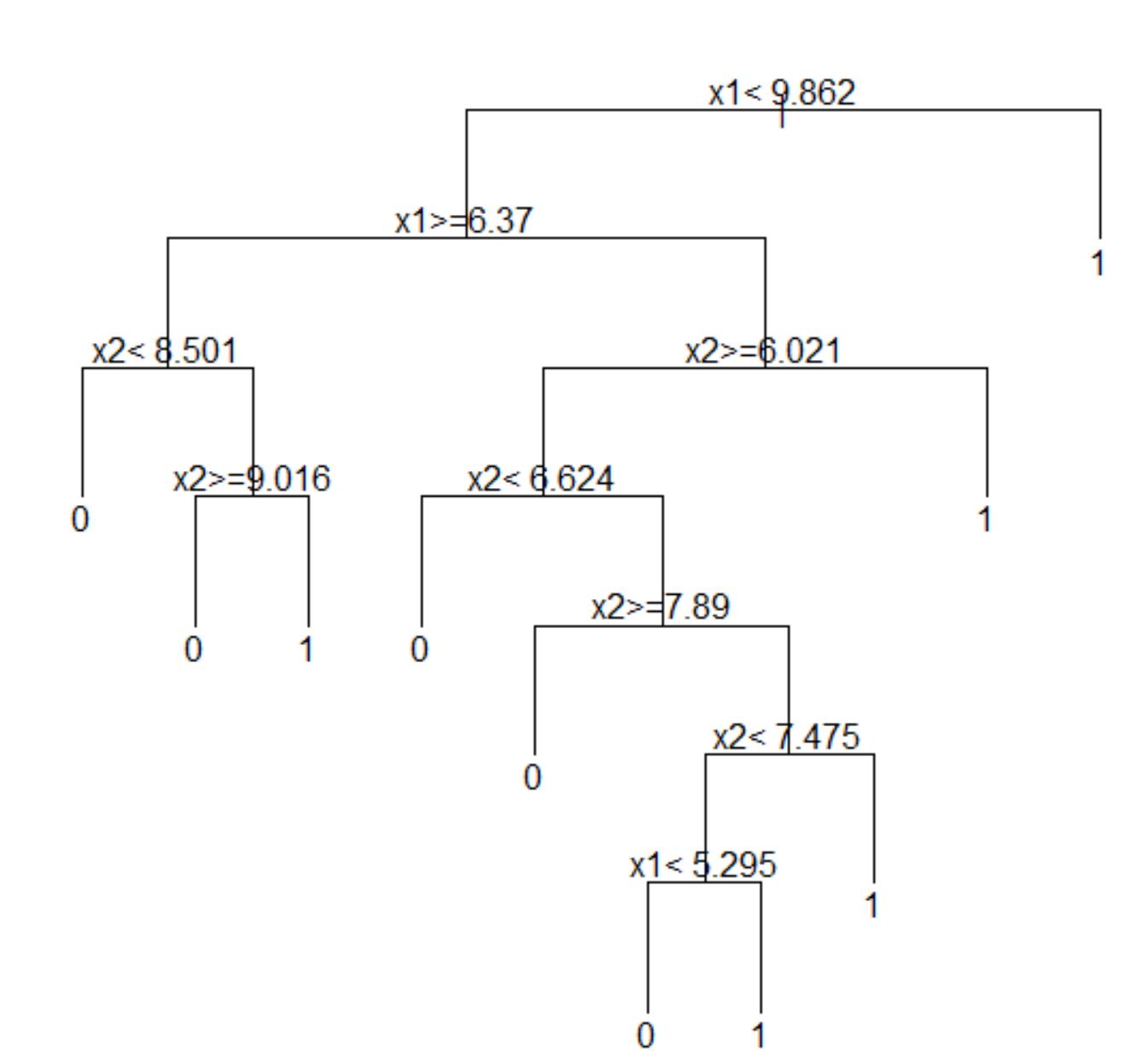}  \\ \\
	\end{tabular}  
	\hskip 2in
		\caption{Classification tree results for Scenario (i)--(iii).}
\end{figure}

\begin{table}[H]\caption{Type I error for our method under the simulation scenarios 1--3 in Section 4.2 of the main paper based on 1000 replications.}	
\begin{center}
\begin{tabular}{  c|c c c  } 
 \hline
Scenario & $\alpha=.01$ & $\alpha=.05$ & $\alpha=.1$ \\\hline
		1 & .013  & .051   & .111   \\   
			2  &  .024  & .060 &  .086     \\  
			3 &  .012 & .049  &  .093 \\  
		 \hline
\end{tabular}
\end{center}
\end{table}

\section{Assumptions}
We state the assumptions that are used for the asymptotic results in Section 3 of the main paper. For simplicity, let $D = (Y,\bm{X},\bm{Z})$, $D_i = (Y_i, \bm{X}_i,\bm{Z}_i)$. We assume the following regularity conditions,
\begin{itemize}
	\item[(C1)] We assume a set of conditions on $f$ for consistency of the MLE. 
	\begin{itemize}
		\item[(a)] $\text{E}|\log f(D; \Psi_0,\bm{\gamma}_0) | < \infty$ for any values of $D$. 
		\item[(b)] There exist compact sets $\Theta$ and $\Gamma$ such that $\text{E}\log \{1 + \sup_{\|\bm{\theta} - \bm{\theta}_0\| \leq \epsilon,\|\bm{\gamma} - \bm{\gamma}_0\| \leq \epsilon }f(D; \bm{\theta},\bm{\gamma})\} < \infty $ for any $\bm{\theta}_0 \in \Theta, \bm{\gamma}_0 \in \Gamma$ and a sufficiently small $\epsilon > 0$.
		\item[(c)] $\text{E}\log \{1 + \sup_{\|\bm{\theta}\|_2 \geq M, \|\bm{\gamma}\|_2 \geq M} f(D; \bm{\theta},\bm{\gamma})\} < \infty$ for a sufficiently large $M$. 
		\item[(d)] $\lim_{\|\bm{\theta}\|_2, \|\bm{\gamma}\|_2 \rightarrow \infty} f(D; \bm{\theta},\bm{\gamma}) =0$ almost surely. 
	\end{itemize} 
	\item[(C2)] Assume that $f$ is four times continuously differentiable with respect to $\bm{\theta}$ and $\bm{\gamma}$.
	\item[(C3)] For any two mixting distributions $\Psi_1$ and $\Psi_2$, if 
	\begin{align*}
	\int f(D; \bm{\theta},\bm{\gamma}) d\Psi_1(\bm{\theta}) = \int f(D; \bm{\theta},\bm{\gamma}) d\Psi_2(\bm{\theta})
	\end{align*}
	for any $D$,	then $\Psi_1 = \Psi_2$. 
	\item[(C4)] There exists a small $\epsilon > 0$ and an integrable function $g$ such that for $h=1,\ldots,m_0$, and every $\bm{\theta}$ satisfying $\| \bm{\theta} - \bm{\theta}_{0h} \| \leq \epsilon$, 
	\begin{align*}
	& |\Delta_{ih} |^3 \leq g(\bm{X}_i),~|u|^3 \leq g(\bm{X}_i),\\
	& |v|^3 \leq g(\bm{X}_i),~\left|\frac{\partial{v}}{\partial \theta_k}\right|^3 \leq g(\bm{X}_i),
	~\left|\frac{\partial{v}^2}{\partial \theta_k \partial \theta_k'}\right|^3 \leq g(\bm{X}_i),
	\end{align*}
	where $u$ is any element in $\bm{Y}_i(\bm{\theta})$, $v$ is any element in $\bm{Z}_i(\bm{\theta})$, and $k,k'$ are indexes from $1,\ldots,p$. 
	\item[(C5)] The covariance matrix of $\bm{b}_i$, denoted by $\bm{B}$, is finite and positive definite. 
	\item[(C6)] The penalty function $p(\beta)$ is continuous and satisfies $p(\beta) \leq 0$ for $\beta \in [0,1]$, $p(.5)=0$, $p(0) = p(1) = -\infty$.  
	\item[(C7)] For every $k=1,\ldots,p$, there exists a constant $C > 0$ and an index $j \in \{1,\ldots,m_0\}$ such that $|\bm{\theta}_{0jk} - \bm{\theta}_{01k} | > C$.
{\color{blue} \item[(C8)] We assume $\frac{\partial^2 f}{\partial \theta_i \theta_j} = o_p(\frac{\partial^2 f}{\partial \theta_i^2})$ for every $i, j=1,\ldots,p$ and $i\neq j$.}
\end{itemize}
To simplify the notation, we drop the subscript $n$ from $\lambda_n$ throughout the proof. Conditions (C1)--(C6) are similar with those in \citet{Li2010} with modifications to accommodate the regression coefficients $\bm{\theta}$ and $\bm{\gamma}$. Condition (C7) essentially rules out the situation where a covariate $x_k$ has the same effect on $y$ in every subgroup. {\color{blue} Condition (C8) essentially assumes the cross-derivatives of the likelihood function $\frac{\partial^2 f}{\partial \theta_i \theta_j}$ are $0$ for $i \neq j$. This is satisfied for a normal linear mixture model where the covariates are uncorrelated with each other, which can hold in our case given that one can first apply a PCA for the covariate space and then perform the usual linear regression on the PC components. For the logistic regression model, this condition is satisfied given covariates are uncorrelated and the predicted individual probabilities $\hat{p}_i (1 - \hat{p}_i) $ are the same for all the patients $i$ in the same subgroup. This condition seems reasonable given that the variability is assumed to be homogeneous for each subgroup. It is unclear how to relax/generalize those assumptions under the current framework, but the simulation results seem to suggest a reasonable level of robustness in our method under the violation of those assumptions. For example, for the logistic mixture model (Section 1.2 of the Web Supplementary File), we find that the type I error is in general very close to the nominal level for most simulation scenarios, although there seems a loss of power especially when $m_0 = 2$ or $3$ due to the violation of this assumption. }

\section{Proof of Theorems 1--3}
\subsection{Proof of Theorem 1}
	The proof extends Theorem 1 of \citet{Li2010} to allow multivariate $\bm{X}$. The main idea is to use mathematical induction. In Lemmas \ref{lm1} and \ref{lm2}, we show that  Theorem 1 holds for $k=1$. In Lemma \ref{lm3}, we show that if Theorem holds for $k=k_0$, then it also holds for $k=k_0+1$. The proof of these lemmas goes to the next section.

\begin{lemma}\label{lm1}
	Suppose the Conditions (C1)--(C3), (C6) and (C7) hold. Under the null distribution $f(D; \Psi_0,\bm{\gamma}_0)$, the following holds for every $\bm{\beta}_0 \in \mathcal{B}^{m_0}$,   
	\begin{align*}
	& \bm{\alpha}^{(1)} - \bm{\alpha}_0 = o_p(1), ~
	  \bm{\beta}^{(1)} - \bm{\beta}_0 = o_p(1),~ \bm{\theta}_1^{(1)} - \bm{\theta}_0 = o_p(1),~ \bm{\theta}_2^{(1)} - \bm{\theta}_0 = o_p(1),~
	\bm{\gamma}^{(1)} - \bm{\gamma}_0 = o_p(1).
	\end{align*}
\end{lemma}

\begin{lemma}\label{lm2}
	Suppose the Conditions (C1)--(C8) hold. Under the null distribution $f(D; \Psi_0,\bm{\gamma}_0)$, the following holds for every $\bm{\beta}_0 \in \mathcal{B}^{m_0}$   ,
	\begin{align*}
	& \bm{\alpha}^{(1)} - \bm{\alpha}_0 = O_p(n^{-1/2}), ~
	 \bm{\beta}^{(1)} - \bm{\beta}_0 = O_p(n^{-1/6}),~
	 \bm{m}_1^{(1)} = O_p(n^{-1/2}), \\
	&\bm{\theta}_1^{(1)} - \bm{\theta}_0 = O_p(n^{-1/4}),~\bm{\theta}_2^{(1)} - \bm{\theta}_0 = O_p(n^{-1/4}), ~ 
	\bm{\gamma}^{(1)} - \bm{\gamma}_0 = O_p(n^{-1/2}).
	\end{align*}
\end{lemma}

\begin{lemma}\label{lm3}
	Suppose that Conditions (C1)--(C8) hold. Under the null distribution $f(D; \Psi_0,\bm{\gamma}_0)$, the following holds for every $\bm{\beta}_0 \in \mathcal{B}^{m_0}$ and some $k \geq 1$, 
	\begin{align*}
	& \bm{\alpha}^{(k)} - \bm{\alpha}_0 = O_p(n^{-1/2}), ~ \bm{\beta}^{(k)} - \bm{\beta}_0 = O_p(n^{-1/6}),~ \bm{m}_1^{(k)} = O_p(n^{-1/2}), \\
	&\bm{\theta}_1^{(k)} - \bm{\theta}_0 = O_p(n^{-1/4}),~
	\bm{\theta}_2^{(k)} - \bm{\theta}_0 = O_p(n^{-1/4}), ~
	\bm{\gamma}^{(k)} - \bm{\gamma}_0 = O_p(n^{-1/2}).
	\end{align*}
	then
		\begin{align*}
		& \bm{\alpha}^{(k+1)} - \bm{\alpha}_0 = O_p(n^{-1/2}), ~ \bm{\beta}^{(k+1)} - \bm{\beta}_0 = O_p(n^{-1/6}),~ \bm{m}_1^{(k+1)} = O_p(n^{-1/2}), \\
		&\bm{\theta}_1^{(k+1)} - \bm{\theta}_0 = O_p(n^{-1/4}),~
		\bm{\theta}_2^{(k+1)} - \bm{\theta}_0 = O_p(n^{-1/4}), ~
		\bm{\gamma}^{(k+1)} - \bm{\gamma}_0 = O_p(n^{-1/2}).
		\end{align*}
\end{lemma}

\subsection{Proof of Theorem 2}
	 We drop $\bm{\gamma}_0$ from $f$ for convenience. The proof contains two parts. We first show that $EM_n^{(K)} \leq  \sup_{\bm{v} \geq 0} \{ 2\bm{v}^T \sum_{i=1}^n \tilde{\bm{b}}_{2i} -n \bm{v}^T \tilde{\bm{B}}_{22}\bm{v} \} +o_p(1)$. Then in the second part, we show the upper bound can be attained. 
	
	(1)	Consider the expansion of $l_n(\hat{\Psi}_0)$ at $\Psi_0$:
	\begin{align*}
	R_{0n} = 2\{l_n (\hat{\Psi}_0) - l_n(\Psi_0)\} = (\sum_{i=1}^n \bm{b}_{1i})^T (n \bm{B}_{11})^{-1} (\sum_{i=1}^n \bm{b}_{1i}) + o_p(1).
	\end{align*}
	Define
	\begin{align*}
	R_{1n} (\Psi^{(k)}(\bm{\beta}_0)) = 2\{pl_n (\Psi^{(k)}(\bm{\beta}_0)) - pl_n^0 (\Psi_0)\} 
	\end{align*}
	By going through the arguments in the proof of Lemma \ref{lm2}, we find that \eqref{eq:e36} holds for $R_{1n} (\Psi^{(k)}(\bm{\beta}_0))$ as well, that is 
	\begin{align}\label{eq:e45}
	R_{1n} (\Psi^{(k)}(\bm{\beta}_0)) \leq 2 \bar{\bm{t}}^T\sum_{i=1}^n \bm{b}_i -n \bar{\bm{t}}^T \bm{B} \bar{\bm{t}} \{1 + o_p(1)\}   + o_p(1),
	\end{align}
	where 
	\begin{align}\label{def:b}
	& \bar{\bm{t}}^T = \left(\alpha_1^{(k)} - \alpha_{01},\ldots,\alpha_{m_0-1}^{(k)} - \alpha_{0m_0-1}; \alpha_1^{(k)} \{\bm{m}_{11}^{(k)}\}^T,\ldots, \alpha_{m_0}^{(k)} \{\bm{m}_{1m_0}^{(k)}\}^T; \alpha_1^{(k)} \{\bm{m}_{21}^{(k)}\}^T,\ldots, \alpha_{m_0}^{(k)} \{\bm{m}_{2m_0}^{(k)}\}^T\right), \nn \\
	& \bm{b}_i = (\bm{b}_{1i}^T, \bm{b}_{2i}^T)^T, ~\bm{b}_{1i} = (\Delta_{i1},\ldots,\Delta_{i m_0-1}, \bm{Y}_i(\bm{\theta}_{01})^T, \ldots, \bm{Y}_i(\bm{\theta}_{0m_0}))^T ,~\bm{b}_{2i} = (\bm{Z}_i(\bm{\theta}_{01})^{T}, \ldots,\bm{Z}_i(\bm{\theta}_{0m_0})^{T})^T .
	\end{align}
where $ \bm{m}_{1h}^{(k)}$ is a $p$-dimensional column vector 
	\begin{align*}
	& \bm{m}_{1h}^{(k)} = \beta_h^{(k)} (\bm{\theta}_{1h}^{(k)} - \bm{\theta}_{0h}) +(1- \beta_h^{(k)}) (\bm{\theta}_{2h}^{(k)} - \bm{\theta}_{0h}),
	\end{align*}
$\bm{m}_{2h}^{(k)}$ is a $p$-dimensional vector with elements
	\begin{align*}
	  \beta_h^{(k)} \left\{\bm{\theta}_{1h}^{(k)}(u) - \bm{\theta}_{0h}(u)\right\}^2 + (1- \beta_h^{(k)}) \left\{\bm{\theta}_{2h}^{(k)}(u) - \bm{\theta}_{0h}(u)\right\}^2 
	\end{align*}
	for $u = 1,\ldots,p$. 
	Recall 
	\begin{align*}
	& \Delta_{ih} = \frac{ f(D_i; \bm{\theta}_{0h}) -  f(D_i; \bm{\theta}_{0m_0})}{f(D_i; \Psi_0)}, \nn \\
	& \bm{Y}_i(\bm{\theta}) = \frac{f'(D_i;\bm{\theta})}{f(D_i; \Psi_0)} \in \mathbb{R}^{p \times 1}~\text{for any}~\bm{\theta} \in \mathbb{R}^{p \times 1}\nn \\
	& \bm{Z}_i(\bm{\theta}) = \frac{1}{f(D_i; \Psi_0)} \times  
   \left( \frac{\partial^2 f(D_i;\bm{\theta})}{\partial \theta_1 \partial \theta_1}, \ldots, \frac{\partial^2 f(D_i;\bm{\theta})}{\partial \theta_p \partial \theta_p}  \right)^T \in \mathbb{R}^{   p    \times 1}.  
	\end{align*} Then due to (C8),
	\begin{align}
	\delta_i = \sum_{h=1}^{m_0-1} (\alpha_h^{(1)}-\alpha_{0h}) \Delta_{ih} + \sum_{h=1}^{m_0} \left\{\alpha_h^{(1)} \{\bm{m}_{1h}^{(1)}\}^T \bm{Y}_i(\bm{\theta}_{0h})  +  \alpha_h^{(1)} \{\bm{m}_{2h}^{(1)} \}^T  \bm{Z}_i(\bm{\theta}_{0h}) \right\} + \sum_{i=1}^{m_0} \epsilon_{ih}. \nn
	\end{align}

Here $\bar{\bm{t}}^T$ is a $(2m_0 p + m_0 - 1)$-dimensional vector, $\bm{b}_{1i}$ is $(m_0 p + m_0 - 1)$-dimensional, $\bm{b}_{2i}$ is $m_0 p$-dimensional.
	Since $\sum_{i=1}^n \bm{b}_i = O_p(n^{1/2})$, we have 
	\begin{align}\label{eq:e451}
	R_{1n} (\Psi^{(k)}(\bm{\beta}_0)) \leq 2 \bar{\bm{t}}^T\sum_{i=1}^n \bm{b}_i -n \bar{\bm{t}}^T \bm{B} \bar{\bm{t}} \{1 + o_p(1)\} + o_p(1),
	\end{align}
    Let $\bm{t} = (\bm{t}_1^T,\bm{t}_2^T)^T$ with
	\begin{align*}
	& \bm{t}_1 = \left(\alpha_1  - \alpha_{01},\ldots,\alpha_{m_0-1}  - \alpha_{0m_0-1}; \alpha_1  \{\bm{m}_{11} \}^T,\ldots, \alpha_{m_0}  \{\bm{m}_{1m_0} \}^T \right)^T, \\
	& \bm{t}_2 = \left( \alpha_1  \bm{m}_{21}^T,\ldots, \alpha_{m_0}  \bm{m}_{2m_0}^T \right)^T.
	\end{align*}
	Then we replace $\bar{\bm{t}}$ in \eqref{eq:e451} by an arbitrary $\bm{t}$ and take the supremum, the following holds,
	\begin{align*}
	R_{1n} (\Psi^{(k)}(\bm{\beta}_0))  - R_{0n} \leq \sup_{\bm{t}} \left\{2 \bm{t}^T\sum_{i=1}^n \bm{b}_i -n \bm{t}^T \bm{B} \bm{t} \right\} - (\sum_{i=1}^n \bm{b}_{1i})^T (n \bm{B}_{11})^{-1} (\sum_{i=1}^n \bm{b}_{1i}) +  o_p(1).
	\end{align*}
	Let $\tilde{\bm{t}}_1 =\bm{t}_1 + \bm{B}_{11}^{-1} \bm{B}_{12} \bm{t}_2$, then because of the orthogonality, 
	\begin{align*}
	2 \bm{t}^T\sum_{i=1}^n \bm{b}_i -n \bm{t}^T \bm{B} \bm{t} = 2 \tilde{\bm{t}}_1^T\sum_{i=1}^n \bm{b}_{1i} -n \tilde{\bm{t}}_1^T \bm{B}_{11} \tilde{\bm{t}}_1 + 2 \tilde{\bm{t}}_2^T\sum_{i=1}^n \bm{b}_{2i} -n \tilde{\bm{t}}_2^T \bm{B}_{22} \tilde{\bm{t}}_2.
	\end{align*}
	Then 
	\begin{align*}
	EM_n^{(k)} \leq & \sup_{\tilde{\bm{t}}_1} \left\{ 2 \tilde{\bm{t}}_1^T\sum_{i=1}^n \bm{b}_{1i} -n \tilde{\bm{t}}_1^T \bm{B}_{11} \tilde{\bm{t}}_1\right\} + \sup_{\tilde{\bm{t}}_2} \left\{  2 \tilde{\bm{t}}_2^T\sum_{i=1}^n \bm{b}_{2i} -n \tilde{\bm{t}}_2^T \bm{B}_{22} \tilde{\bm{t}}_2\right\} \nn \\
	& ~~~ - \left(\sum_{i=1}^n \bm{b}_{1i}\right)^T (n \bm{B}_{11})^{-1} (\sum_{i=1}^n \bm{b}_{1i}) +  o_p(1) \\
	& = \left(\sum_{i=1}^n \bm{b}_{1i}\right)^T (n \bm{B}_{11})^{-1} (\sum_{i=1}^n \bm{b}_{1i}) + \sup_{\tilde{\bm{t}}_2} \left\{  2 \tilde{\bm{t}}_2^T\sum_{i=1}^n \bm{b}_{2i} -n \tilde{\bm{t}}_2^T \bm{B}_{22} \tilde{\bm{t}}_2\right\} \nn \\
	& ~~~ - \left(\sum_{i=1}^n \bm{b}_{1i}\right)^T (n \bm{B}_{11})^{-1} (\sum_{i=1}^n \bm{b}_{1i}) +  o_p(1) \\
	& = \sup_{\tilde{\bm{t}}_2} \left\{  2 \tilde{\bm{t}}_2^T\sum_{i=1}^n \bm{b}_{2i} -n \tilde{\bm{t}}_2^T \bm{B}_{22} \tilde{\bm{t}}_2\right\} +  o_p(1).
	\end{align*}
	Note that all the elements of $\bm{t}_2$ are non-negative, this proves the first part. 
	
	(2) Next we show that the upper bound can be attained. Since EM algorithm always increases the $pl_n$, it is good enough to consider $k=1$. In the definition of $\bar{\bm{t}}$, there are a total of free $ \left\{2m_p p + m_0 -1\right\}$ parameters. Hence we can find $\alpha_h$, $\bm{\theta}_{1h},\bm{\theta}_{2h}$ such that the corresponding $\bar{\bm{t}}_2$ attained the supremum in $\sup_{\tilde{\bm{t}}_2} \left\{  2 \tilde{\bm{t}}_2^T\sum_{i=1}^n \bm{b}_{2i} -n \tilde{\bm{t}}_2^T \bm{B}_{22} \tilde{\bm{t}}_2\right\}$. Let $
	\bar{\bm{t}}_1= (n\bm{B}_{11})^{-1} \sum_{i=1}^n \bm{b}_{1i} - \bm{B}_{11}^{-1} \bm{B}_{12} $. Then the resulting $EM_n^{(1)}$ attains the upper bound, hence concludes the proof.

\subsection{Proof of Theorem 3}
	Let $\hat{\bm{v}} = \text{argmax}_{\bm{v} \geq 0} \{ 2\bm{v}^T \sum_{i=1}^n \tilde{\bm{b}}_{2i} -n \bm{v}^T \tilde{\bm{B}}_{22}\bm{v} \} $. By Lemma 6 of \citet{Li2010}, which is first given by \citet{Kudo1963}, we have $(\bm{w}^T - \hat{\bm{v}}^T \tilde{\bm{B}}_{22}) \hat{\bm{v}} = 0$. Therefore
	\begin{align}
	\sup_{\bm{v} \geq 0} (2\bm{v}^T \bm{w} - \bm{v}^T \tilde{\bm{B}}_{22}\bm{v} ) =  \hat{\bm{v}}^T \tilde{\bm{B}}_{22} \hat{\bm{v}}.
	\end{align}
	Now we only need to find out the distribution of $\hat{\bm{v}}^T \tilde{\bm{B}}_{22} \hat{\bm{v}}$. Note $\hat{\bm{v}}$ has $ m_0 p   $ elements. Depending on the positiveness of these elements, we could form $2^{m_0 p }$ regions for $\bm{w}$. More precisely, let $\hat{\bm{v}} = (\hat{\bm{v}}_1,\hat{\bm{v}}_2)^T$ be a partition of $\hat{\bm{v}}$, in which $\hat{\bm{v}}_1$ is a vector of zeros of dimension $\{m_0 p   -s\}$, and $ \hat{\bm{v}}_2$ is a vector of rest positive elements of dimension $s$. We can partition $\tilde{\bm{B}}_{22}$ correspondingly as
	\begin{align*}
	\tilde{\bm{B}}_{22} = \left(\begin{array}{cc}
	\bm{W}_{11} & \bm{W}_{12} \\
	\bm{W}_{21} & \bm{W}_{22} \\
	\end{array}\right).
	\end{align*}
	By Lemma 6 of \citet{Li2010}, $\hat{\bm{v}}_1=0$ and $\hat{\bm{v}}_2 > 0$ is equivalent with
	\begin{align*}
	\bm{W}_{22}^{-1} \bm{w}_2 > 0, ~~~ \bm{W}_{12} \bm{W}_{22}^{-1} \bm{w}_2  - \bm{w}_1 > 0.
	\end{align*}
	Therefore the partition $\hat{\bm{v}} = (\hat{\bm{v}}_1,\hat{\bm{v}}_2)^T$ is uniquely related with the region of $\bm{w}$, which is defined by
	\begin{align*}
	\mathcal{U} = \{\bm{w}: \bm{W}_{22}^{-1} \bm{w}_2 > 0,~ \bm{W}_{12} \bm{W}_{22}^{-1} \bm{w}_2  - \bm{w}_1 > 0\}.
	\end{align*}
	Denote the collection of all possible $\mathcal{U}$ by $\mathcal{S}$, then for any $x$, 
	\begin{align*}
	\text{P}(\hat{\bm{v}}^T \tilde{\bm{B}}_{22} \hat{\bm{v}} \leq x) = \sum_{\mathcal{U} \in \mathcal{S}}  \text{P}(
	\hat{\bm{v}}^T \tilde{\bm{B}}_{22} \hat{\bm{v}} \leq x | \bm{w} \in \mathcal{U} ) \text{P}(\bm{w} \in \mathcal{U}).
	\end{align*}
	Because $\hat{\bm{v}}^T \tilde{\bm{B}}_{22} \hat{\bm{v}} = \bm{w}_2^T \bm{W}_{22}^{-1} \bm{w}_2$
	\begin{align*}
	\text{P}(\hat{\bm{v}}^T \tilde{\bm{B}}_{22} \hat{\bm{v}} \leq x | \bm{w} \in \mathcal{U} ) & = \text{P}(\bm{w}_2^T \bm{W}_{22}^{-1} \bm{w}_2 \leq x | \bm{W}_{22}^{-1} \bm{w}_2 > 0，~ \bm{W}_{12} \bm{W}_{22}^{-1} \bm{w}_2  - \bm{w}_1 > 0) \\
	& = \text{P}(\bm{w}_2^T \bm{W}_{22}^{-1} \bm{w}_2 \leq x |  \bm{W}_{22}^{-1} \bm{w}_2 > 0)
	\end{align*}
	because of the normality of $\bm{w}$, and the fact that $\bm{w}_2$ and $\bm{W}_{12} \bm{W}_{22}^{-1} \bm{w}_2  - \bm{w}_1 $ are independent. Moreover, $\bm{w}_2^T \bm{W}_{22}^{-1} \bm{w}_2$ and $\bm{W}_{22}^{-1/2} \bm{w}_2 / \sqrt{\bm{w}_2^T \bm{W}_{22}^{-1} \bm{w}_2}$ are independent. Hence
	\begin{align*}
	\text{P}(\hat{\bm{v}}^T \tilde{\bm{B}}_{22} \hat{\bm{v}} \leq x | \bm{w} \in \mathcal{U} ) 
	& = \text{P}(\bm{w}_2^T \bm{W}_{22}^{-1} \bm{w}_2 \leq x | \bm{W}_{22}^{-1/2} \bm{w}_2  / \sqrt{\bm{w}_2^T \bm{W}_{22}^{-1} \bm{w}_2}> 0) \\
	& = \text{P}(\bm{w}_2^T \bm{W}_{22}^{-1} \bm{w}_2 \leq x).
	\end{align*}
	In other words, $\hat{\bm{v}}^T \tilde{\bm{B}}_{22} \hat{\bm{v}}$ follows $\chi_{s}^2$ distribution where $s$ is the number of positive elements in $\hat{\bm{v}}$. Let $\mathcal{U}_s$ be the collection of $\bm{w}$ whose corresponding $\hat{\bm{v}}$ has $s$ positive components. Define $a_s = P(\mathcal{U}_s)$ and $\bar{s} = m_0 p $, then
	$
	\hat{\bm{v}}^T \tilde{\bm{B}}_{22} \hat{\bm{v}} \sim \sum_{s=0}^{\bar{s}} a_s \chi_s^2
	$, the conclusion follows.

\section{Proof of Lemmas 1--3}
\subsection{Proof of Lemma \ref{lm1}}
	Note that $\hat{\Psi}_0 \in \Omega_{2m_0}(\bm{\beta}_0)$. Hence $\Psi^{(1)} \in \Omega_{2m_0}(\bm{\beta}_0)$ and it maximizes $pl_n(\Psi)$ within the class. Therefore $pl_n(\Psi^{(1)}) \geq pl_n(\hat{\Psi}_0)$. Hence $$l_n(\Psi^{(1)})  + p(\bm{\beta}_0)   = pl_n(\Psi^{(1)})\geq pl_n(\hat{\Psi}_0) \geq l_n(\Psi_0) + p(\bm{\beta}_0), $$ which gives $l_n(\Psi^{(1)}) \geq    l_n(\Psi_0) $. Following the same arguments of the consistency of MLE in the proof of Lemma 1 in \citet{Li2010}, we conclude that $\|\Psi^{(1)}(\bm{\beta}_0) - \Psi_0\| \rightarrow 0$ almost surely with $\|F(\bm{\theta})\|=\sup_{\bm{\theta} \in \Theta} |F(\bm{\theta})|$. Then we have the consistency of $\alpha^{(1)}$, $\bm{\beta}^{(1)}$. Given the consistency of $\Psi$, We also have the consistency of $f(D;\bm{\theta}^{(1)},\bm{\gamma}^{(1)})$, hence the consistency of $\bm{\theta}_1^{(1)}$, $ \bm{\theta}_2^{(1)}$ and $\bm{\gamma}^{(1)}$ holds as well.

\subsection{Proof of Lemma \ref{lm2}}
Let $\{l_n(\Psi^{(1)}(\bm{\beta}_0)) - l_n(\Psi_0)\}=  \sum_{i=1}^n \log(1+\delta_i)$, 
	where
	\begin{align}
	\delta_i & = \frac{f(D_i; \Psi^{(1)}(\bm{\beta}_0)) - f(D_i; \Psi_0)}{ f(D_i; \Psi_0)} \nn \\
	& = \sum_{h=1}^{m_0} (\alpha_h^{(1)}-\alpha_{0h}) \frac{f(Di; \bm{\theta}_{0h})}{f(D_i; \Psi_0)} \nn \\
	&~~~~~~     + \sum_{h=1}^{m_0} \alpha_h^{(1)} \left\{\beta_h^{(1)}  \frac{f(D_i; \bm{\theta}_{1h}^{(1)}) - f(D_i;\bm{\theta}_{0h}) }{f(D_i;\Psi_0)} + (1-\beta_h^{(1)})  \frac{f(D_i; \bm{\theta}_{2h}^{(1)}) - f(D_i; \bm{\theta}_{0h}) }{f(D_i; \Psi_0)}  \right\} \nn \\
	&= \sum_{h=1}^{m_0-1} (\alpha_h^{(1)}-\alpha_{0h}) \frac{f(D_i; \bm{\theta}_{0h}) - f(D_i; \bm{\theta}_{0m_0})}{f(D_i; \Psi_0)} \nn \\
	&~~~~~~     + \sum_{h=1}^{m_0} \alpha_h^{(1)} \left\{\beta_h^{(1)}  \frac{f(D_i; \bm{\theta}_{1h}^{(1)}) - f(D_i;\bm{\theta}_{0h}) }{f(D_i; \Psi_0)} + (1-\beta_h^{(1)})  \frac{f(D_i; \bm{\theta}_{2h}^{(1)}) - f(D_i; \bm{\theta}_{0h}) }{f(D_i;\Psi_0)}  \right\}.
	\end{align}
	Note that
	\begin{align*}
	f(D_i; \bm{\theta}_{1h}^{(1)}) - f(D_i;\bm{\theta}_{0h}) & = (\bm{\theta}_{1h}^{(1)} - \bm{\theta}_{0h})^T f'(D_i; \bm{\theta}_{0h}) + \frac{1}{2} (\bm{\theta}_{1h}^{(1)} - \bm{\theta}_{0h})^T f''(D_i; \bm{\theta}_{0h})  (\bm{\theta}_{1h}^{(1)} - \bm{\theta}_{0h}) + \epsilon_{ih}.
	\end{align*}
	Define
	\begin{align}
	& \bm{m}_{1h}^{(k)} = \beta_h^{(k)} (\bm{\theta}_{1h}^{(k)} - \bm{\theta}_{0h}) +(1- \beta_h^{(k)}) (\bm{\theta}_{2h}^{(k)} - \bm{\theta}_{0h}).
	\end{align}
	For index $u = 1,\ldots,p$, let $\bm{m}_{2h}^{(k)}$ be a $p$-dimensional vector with elements
	\begin{align*}
	{\color{blue} \beta_h^{(k)} \left\{\bm{\theta}_{1h}^{(k)}(u) - \bm{\theta}_{0h}(u)\right\}^2 + (1- \beta_h^{(k)}) \left\{\bm{\theta}_{2h}^{(k)}(u) - \bm{\theta}_{0h}(u)\right\}^2. }
	\end{align*}
	Let $\bm{m}_1^{(k)} = (\bm{m}_{11}^{(k)},\ldots,\bm{m}_{1m_0}^{(k)})$ and $\bm{m}_2^{(k)} = (\bm{m}_{21}^{(k)},\ldots,\bm{m}_{2m_0}^{(k)})$ be two matrices. Define
	\begin{align}
	& \Delta_{ih} = \frac{ f(D_i; \bm{\theta}_{0h}) -  f(D_i; \bm{\theta}_{0m_0})}{f(D_i; \Psi_0)}, \nn \\
	& \bm{Y}_i(\bm{\theta}) = \frac{f'(D_i;\bm{\theta})}{f(D_i; \Psi_0)} \in \mathbb{R}^{p \times 1}~\text{for any}~\bm{\theta} \in \mathbb{R}^{p \times 1}\nn \\
	& \bm{Z}_i(\bm{\theta}) = \frac{1}{f(D_i; \Psi_0)} \times  
 {\color{blue} \left( \frac{\partial^2 f(D_i;\bm{\theta})}{\partial \theta_1 \partial \theta_1}, \ldots, \frac{\partial^2 f(D_i;\bm{\theta})}{\partial \theta_p \partial \theta_p}  \right)^T \in \mathbb{R}^{   p    \times 1},} \nn \\
	& \bm{b}_{1i} = (\Delta_{i1},\ldots,\Delta_{i m_0-1}, \bm{Y}_i(\bm{\theta})^T, \ldots, \bm{Y}_i(\bm{\theta})^T)^T \in \mathbb{R}^{(m_0-1 + m_0 p) \times 1}, \nn \\
	& \bm{b}_{2i} = (\bm{Z}_i(\bm{\theta}_{01})^{T}, \ldots,\bm{Z}_i(\bm{\theta}_{0m_0})^{T})^T \in \mathbb{R}^{ m_0 p    \times 1},
	\end{align}
	and $\bm{b}_i = (\bm{b}_{1i}^T, \bm{b}_{2i}^T)^T$. Then 
	\begin{align}
	\delta_i = \sum_{h=1}^{m_0-1} (\alpha_h^{(1)}-\alpha_{0h}) \Delta_{ih} + \sum_{h=1}^{m_0} \left\{\alpha_h^{(1)} \{\bm{m}_{1h}^{(1)}\}^T \bm{Y}_i(\bm{\theta}_{0h})  +  \alpha_h^{(1)} \{\bm{m}_{2h}^{(1)} \}^T  \bm{Z}_i(\bm{\theta}_{0h}) \right\} + \sum_{i=1}^{m_0} \epsilon_{ih}. \nn
	\end{align}
	For $i=1,\ldots,n$ and index $u$, we define a $p$-dimensional vector $\bm{U}_{ih}(\bm{\theta})$ as the collection of elements
	\begin{align*}
	\frac{f(D_i; \bm{\theta}) - f(D_i; \bm{\theta}_{0h})-f'(D_i; \bm{\theta}_{0h})(\bm{\theta}-\bm{\theta}_{0h})-
		(\bm{\theta}-\bm{\theta}_{0h})^T f''(D_i; \bm{\theta}_{0h})  (\bm{\theta}-\bm{\theta}_{0h})/2 }{f(D_i; \Psi_0) \left\{\bm{\theta}(u) -\bm{\theta}_{0h}(u)\right\}^3},
	\end{align*}
	where $\bm{\theta}(u)$ is the $u$-th element of $\bm{\theta}$. Then Condition (C4) implies that for every $h=1,\ldots,m_0$, the process $n^{-1/2} \sum_{i=1}^n \bm{U}_{ih} (\bm{\theta})$ is tight for any $\bm{\theta} \in B(\bm{\theta}_{0h}, \epsilon_0)$ \citep{Billing1968}. Then 
	\begin{align}
	n^{-1/2} \sum_{i=1}^{n} \bm{U}_{ih}(\tilde{\bm{\theta}}) = O_p(1)
	\end{align}
	when $\tilde{\bm{\theta}}$ is a consistent estimator of $\bm{\theta}$. 
	
	Again, for index $u  = 1,\ldots,p$, define $\bm{\theta}^3$ as a $p$-dimensional vector with elements of $\bm{\theta}(u)^3$. Then
	\begin{align}
	\epsilon_{ih} = \alpha_h^{(1)}  \left\{ \beta_h^{(1)}  \{(\bm{\theta}_{1h}^{(1)} -\bm{\theta}_{0h})^3\}^T \bm{U}_{ih}(\bm{\theta}_{1h}^{(1)})    + (1- \beta_h^{(1)})  \{(\bm{\theta}_{2h}^{(1)} - \bm{\theta}_{0h})^3\}^T \bm{U}_{ih}(\bm{\theta}_{2h}^{(1)})   \right\}.
	\end{align}
	Then 
	\begin{align}
	\sum_{i=1}^n \delta_i = \sum_{i=1}^n \left\{ \sum_{h=1}^{m_0-1} (\alpha_h^{(1)}-\alpha_{0h}) \Delta_{ih} + \sum_{h=1}^{m_0} \left\{\alpha_h^{(1)} \{\bm{m}_{1h}^{(1)}\}^T \bm{Y}_i(\bm{\theta}_{0h})  +  \alpha_h^{(1)} \{\bm{m}_{2h}^{(1)} \}^T  \bm{Z}_i(\bm{\theta}_{0h}) \right\}  \right\} + \epsilon_n, \nn
	\end{align}
	where $|\epsilon_n |$ equals to
	\begin{align}
	& \left|\sum_{i=1}^n \sum_{h=1}^{m_0} \epsilon_{ih} \right| \nn \\
	&\leq  n^{1/2} \sum_{h=1}^{m_0} \alpha_h^{(1)} \left\{ \beta_h^{(1)}  \{|\bm{\theta}_{1h}^{(1)} -\bm{\theta}_{0h}|^3\}^T n^{-1/2} \sum_{i=1}^n \bm{U}_{ih} (\bm{\theta}_{1h}^{(1)}) + (1- \beta_h^{(1)})  \{|\bm{\theta}_{2h}^{(1)} - \bm{\theta}_{0h}|^3\}^T n^{-1/2} \sum_{i=1}^n \bm{U}_{ih}(\bm{\theta}_{2h}^{(1)})    \right\} \nn \\
	&= O_p(n^{1/2}) \sum_{h=1}^{m_0} \alpha_h^{(1)} \left\{ \beta_h^{(1)}  \{|\bm{\theta}_{1h}^{(1)} -\bm{\theta}_{0h}|^3\}^T +  (1- \beta_h^{(1)})  \{|\bm{\theta}_{2h}^{(1)} - \bm{\theta}_{0h}|^3\}^T \right\} \bm{1} \nn \\
	& = o_p(n^{1/2}) \sum_{h=1}^{m_0} \{\bm{m}_{2h}^{(1)} \}^T\bm{1} \nn \\
	& \leq o_p(1) +o_p(n) \sum_{h=1}^{m_0} \| \bm{m}_{2h}^{(1)} \|_{\infty}^2.
	\end{align}
	Let 
	\begin{align*}
	\bar{\bm{t}} = \left(\alpha_1^{(1)} - \alpha_{01},\ldots,\alpha_{m_0-1}^{(1)} - \alpha_{0m_0-1}; \alpha_1^{(1)} \{\bm{m}_{11}^{(1)}\}^T,\ldots, \alpha_{m_0}^{(1)} \{\bm{m}_{1m_0}^{(1)}\}^T; \alpha_1^{(1)} \{\bm{m}_{21}^{(1)}\}^T,\ldots, \alpha_{m_0}^{(1)} \{\bm{m}_{2m_0}^{(1)}\}^T\right)^T
	\end{align*}
	and $\bm{b}_i = (\bm{b}_{1i}^T,\bm{b}_{2i}^T )^T$. 
	Then 
	\begin{align}
	\sum_{i=1}^n \delta_i = \sum_{i=1}^n \bar{\bm{t}}^T \bm{b}_i + \epsilon_n.
	\end{align}
	Using the fact that $\log (1+x) \leq x - x^2/2 + x^3/3$, we have
	\begin{align}
	 2\sum_{i=1}^n \log (1+\delta_i) 
	& \leq 2 \sum_{i=1}^n \delta_i  - \sum_{i=1}^n \delta_i^2 + \frac{2}{3} \sum_{i=1}^n \delta_i^3 \nn \\
	&          = 2  \sum_{i=1}^n \bar{\bm{t}}^T \bm{b}_i -  \sum_{i=1}^n (\bar{\bm{t}}^T \bm{b}_i)^2 + \frac{2}{3} \sum_{i=1}^n (\bar{\bm{t}}^T \bm{b}_i)^3 + O_p(\epsilon_n). 
	\end{align}
	By Condition (C5), 
	\begin{align}
	& \sum_{i=1}^n (\bar{\bm{t}}^T \bm{b}_i)^2 = n\bar{\bm{t}}^T \bm{B} \bar{\bm{t}}\{1 + o_p(1)\}, \nn \\
	& \sum_{i=1}^n (\bar{\bm{t}}^T \bm{b}_i)^3 =o_p(n) \bar{\bm{t}}^T\bar{\bm{t}}.
	\end{align}
	Note that $\epsilon_n = o_p(1) + o_p(n) \bar{\bm{t}}^T	\bar{\bm{t}}$, hence
	\begin{align}\label{eq:e36}
	2\sum_{i=1}^n \log (1+\delta_i)  \leq 2 \bar{\bm{t}}^T\sum_{i=1}^n \bm{b}_i -n \bar{\bm{t}}^T \bm{B} \bar{\bm{t}} \{1 + o_p(1)\} + o_p(1).
	\end{align}
Because $pl_n(\Psi^{(1)}) \geq pl_n(\Psi_0)$, we have	
	\begin{align}\label{eq:e222}
	2 \{ l_n(\Psi^{(1)}) - l_n(\Psi_0) \} + 2 n\left\{q_{\lambda}(\Psi_0)- q_{\lambda}(\Psi^{(1)}) \right\} \geq 2 p(\bm{\beta_0}).
	 \end{align}
		Observe that 
		\begin{align}\label{eq:qq}
		& | q_{\lambda}(\Psi_0) -q_{\lambda}(\Psi^{(1)}(\bm{\beta}_0)) | \leq \lambda \times \nn \\
		&	 \sum_{k=1}^p \frac{ |\sum_{j=2}^{m_0} (\theta_{0jk} - \theta_{01k})^2 + \sum_{j'=1}^{m_0} (\theta_{0j'k} - \theta_{01k})^2- \sum_{j=2}^{m_0} (\theta_{1jk} - \theta_{11k})^2 - \sum_{j'=1}^{m_0} (\theta_{2j'k} - \theta_{11k})^2 |}{ \{ \sum_{j=2}^{m_0} (\theta_{0jk} - \theta_{01k})^2 + \sum_{j'=1}^{m_0} (\theta_{0j'k} - \theta_{01k})^2\}^{1/2} + \{ \sum_{j=2}^{m_0} (\theta_{1jk} - \theta_{11k})^2 + \sum_{j'=1}^{m_0} (\theta_{2j'k} - \theta_{11k})^2\}^{1/2}  } \nn \\
		& \leq C_1 \lambda \left|\sum_{j=2}^{m_0} (\theta_{0jk} - \theta_{01k})^2 + \sum_{j'=1}^{m_0} (\theta_{0j'k} - \theta_{01k})^2- \sum_{j=2}^{m_0} (\theta_{1jk} - \theta_{11k})^2 - \sum_{j'=1}^{m_0} (\theta_{2j'k} - \theta_{11k})^2 \right| \nn \\
		& \leq C_2 |\bar{\bm{t}}^T \bm{1}|
		\end{align}
		for some positive constants $C_1,C_2$ because of the Condition (C8). Note that $n \lambda = o(n^{1/2})$ due to Condition (C7), \eqref{eq:e222} and \eqref{eq:e36} together implies
		\begin{align}
		2p(\bm{\beta}_0) \leq 2 \bar{\bm{t}}^T \left\{\sum_{i=1}^n \bm{b}_i +o_p(n^{1/2}) \bm{1}\right\} -n \bar{\bm{t}}^T \bm{B} \bar{\bm{t}} \{1 + o_p(1)\} + o_p(1)
		\end{align}

	Because $n^{-1/2} \sum_{i=1}^n \bm{b}_i = O_p(1)$ and $p(\bm{\beta}_0)$ is fixed, we have $\bar{\bm{t}} = O_p(n^{-1/2})$. This implies $\alpha_1^{(s)} - \alpha_{0s} = O_p(n^{-1/2})$ for $s=1,\ldots,m_0-1$, $\bm{m}_{1t}^{(1)} = O_p(n^{-1/2})$ and $\bm{m}_{2t}^{(1)} = O_p(n^{-1/2})$ for every $t=1,\ldots,m_0$. The $n^{-1/2}$ convergence rate of $\bm{m}_{2t}^{(1)}$ implies the $n^{-1/2}$ convergence rate of $\bm{\theta}_{1h}^{(k)}$ and $\bm{\theta}_{2h}^{(k)}$. For the rate of $\bm{\beta}^{(1)}$, it holds trivially because $\bm{\beta}^{(1)} = \bm{\beta}_0$ by definition. Once we obtain the convergence rate of the parameters in $\Psi^{(1)}$, the convergence rate of $\bm{\gamma}^{(1)}$ is obvious. This completes the proof.

\subsection{Proof of Lemma \ref{lm3}}
	We divide the proof into three parts. 
	
	(1) First, we show $\bm{\beta}^{(k+1)} - \bm{\beta}_0 = O_p(n^{-1/6})$. Here we first proof a useful result,
	\begin{align}\label{eq:lm31}
	\sum_{i=1}^n \frac{f(D_i;\bm{\theta}_{1h}^{(k)})}{f(D_i;\Psi^{(k)}(\bm{\beta}_0))} = n\{1 + O_p(n^{-1/6})\}.
	\end{align}
	Note that for every $j=1,2$ and $h=1,\ldots,p$, 
	\begin{align}
	\frac{f(D_i;\bm{\theta}_{jh}^{(k)}) - f(D_i;\bm{\theta}_{0h})}{f(D_i;\Psi_0)} = \bm{Y}_i(\bm{\theta}_{0h})^T (\bm{\theta}_{jh}^{(k)}-\bm{\theta}_{0h}) + (\bm{\theta}_{jh}^{(k)}-\bm{\theta}_{0h})^T  \tilde{\bm{Z}}_i(\tilde{\bm{\theta}}_{jh}^{(k)}) (\bm{\theta}_{jh}^{(k)}-\bm{\theta}_{0h}), 
	\end{align}
	where $\tilde{\bm{Z}}_i(\tilde{\bm{\theta}}_{jh}^{(k)})$ is a $p$ by $p$ diagnoal matrix defined by
	\begin{align}
	\tilde{\bm{Z}}_i(\bm{\theta}) =\frac{1}{f(D_i; \Psi_0)}  \text{diag} \left\{ 
	\frac{\partial^2 f(D_i;\bm{\theta} )}{\partial \theta_1^2}, \ldots, 	\frac{\partial^2 f(D_i;\bm{\theta} )}{\partial \theta_p^2}\right\},
	\end{align}
	and $\tilde{\bm{\theta}}_{jh}^{(k)} $ is in the neighborhood of $\bm{\theta}_{0h}$. Then
	\begin{align}
	\delta_i &= \frac{f(D_i; \Psi^{(k)}(\bm{\beta}_0)) - f(D_i; \Psi_0)}{f(D_i; \Psi_0)} \nn \\
	&= \sum_{h=1}^{m_0-1} (\alpha_h^{(k)}-\alpha_0) \Delta_{ih} + \sum_{h=1}^{m_0} \alpha_h^{(k)} \{\bm{m}_{1h}^{(k)}\}^T \bm{Y}_i(\bm{\theta}_{0h})  
	+ \sum_{h=1}^{m_0} \alpha_h^{(k)}\beta_h^{(k)} (\bm{\theta}_{1h}^{(k)}-\bm{\theta}_{0h})^T  \tilde{\bm{Z}}_i(\tilde{\bm{\theta}}_{1h}^{(k)}) (\bm{\theta}_{1h}^{(k)}-\bm{\theta}_{0h}) \nn \\ & ~~ +
	\sum_{h=1}^{m_0} \alpha_h^{(k)}\beta_h^{(k)} (\bm{\theta}_{2h}^{(k)}-\bm{\theta}_{0h})^T  \tilde{\bm{Z}}_i(\tilde{\bm{\theta}}_{2h}^{(k)}) (\bm{\theta}_{2h}^{(k)}-\bm{\theta}_{0h}). \nn
	\end{align}
	Because $\Delta_{ih}$ has a constant upper bound, $\max_{1\leq i \leq n} \sum_{h=1}^{m_0-1} |\alpha_h^{(k)}-\alpha_0) \Delta_{ih}| = O_p(n^{-1/4})$. For the second term of $\delta_i$, because of Condition (C4), $\max_{1\leq i \leq n} \|\bm{Y}_i(\bm{\theta}_{0h})\|_{\infty} = O_p(n^{1/3})$. From the proof of Lemma \ref{lm2}, $\bm{m}_{1h}^{(k)} = O_p(n^{-1/2})$. Therefore the second term of $\delta_i$ is $O_p(n^{-1/6})$ uniformly. For the third and fourth terms, Condition (C4) implies $\max_{1\leq i \leq n} \|\tilde{\bm{Z}}_i(\tilde{\bm{\theta}}_{2h}^{(k)})\|_2 = O_p(n^{1/3})$. Combine with the assumption that $\bm{\theta}_j^{(k)} - \bm{\theta}_0 = O_p(n^{-1/4})$ for $j=1,2$, both third and fourth terms of $\delta_i$ are $O_p(n^{-1/6})$. Therefore $\max_{1\leq i \leq n} |\delta_i| = O_p(n^{-1/6})$. In other words, uniformly in $i$, 
	\begin{align*}
	&\frac{1}{n} \sum_{i=1}^n \frac{f(D_i;\bm{\theta}_{1h}^{(k)})}{f(D_i;\Psi^{(k)}(\bm{\beta}_0)} \nn \\
	& = 
	\frac{1}{n} \sum_{i=1}^n \frac{f(D_i;\bm{\theta}_{1h}^{(k)})}{f(D_i;\Psi_0)} 
	\frac{f(D_i;\Psi_0)}{f(D_i;\Psi^{(k)}(\bm{\beta}_0)} \nn \\
	& = \frac{1}{n} \sum_{i=1}^n \frac{f(D_i;\bm{\theta}_{1h}^{(k)})}{f(D_i;\Psi_0)} \{1+O_p(n^{-1/6})\} \nn \\
	& = \frac{1}{n} \sum_{i=1}^n \left\{\frac{f(D_i;\bm{\theta}_{0h})}{f(D_i;\Psi_0)}   + \bm{Y}_i(\bm{\theta}_{0h})^T (\bm{\theta}_{jh}^{(k)}-\bm{\theta}_{0h}) + (\bm{\theta}_{jh}^{(k)}-\bm{\theta}_{0h})^T  \tilde{\bm{Z}}_i(\tilde{\bm{\theta}}_{jh}^{(k)}) (\bm{\theta}_{jh}^{(k)}-\bm{\theta}_{0h})\right\}  \{1+O_p(n^{-1/6})\} \nn \\
	& = 1+O_p(n^{-1/6}). 
	\end{align*}
	This shows \eqref{eq:lm31}. By the definition, $\beta_h^{(k+1)}$ maximizes 
	\begin{align}
	Q_{nh}(\beta) = \sum _{i=1}^n w_{i1h}^{(k)} \log(\beta) + \sum _{i=1}^n w_{i2h}^{(k)} \log(1 - \beta) + p(\beta) \nn 
	\end{align}
	with 
	\begin{align}
	&  \sum _{i=1}^n w_{i1h}^{(k)} =\sum _{i=1}^n \alpha_h^{(k)} \beta_h^{(k)} \frac{f(D_i;\bm{\theta}_{1h}^{(k)})}{f(D_i;\Psi^{(k)}(\bm{\beta}_0)} = n \alpha_h^{(k)} \beta_h^{(k)}  \{1+O_p(n^{-1/6})\}, \nn \\
	& \sum _{i=1}^n w_{i2h}^{(k)} = n \alpha_h^{(k)} - \sum _{i=1}^n w_{i1h}^{(k)} = n \alpha_h^{(k)} (1- \beta_h^{(k)})  \{1+O_p(n^{-1/6})\}
	\end{align}
	Let 
	\begin{align}
	H_{nh}(\beta) = \sum _{i=1}^n w_{i1h}^{(k)} \log(\beta) + \sum _{i=1}^n w_{i2h}^{(k)} \log(1 - \beta).
	\end{align}
	Its maximizer is
	\begin{align}
	\hat{\beta}_h = \beta_h^{(k)} \{1+O_p(n^{-1/6})\}.
	\end{align}
	Then consider any $\beta^*$ such that $n^{-1/6} \leq |\beta^* - \hat{\beta}_h | \leq n^{-1/3}$, by Taylor expansion, 
	\begin{align*}
	H_{nh}(\hat{\beta}_h) - H_{nh}(\beta^*) \geq \epsilon \alpha_h^{(k)} n (\beta^* - \hat{\beta}_h )^2 \geq \epsilon \alpha_h^{(k)} n^{2/3}
	\end{align*}
	for some $\epsilon > 0$. Then 
	\begin{align*}
	Q_{nh}(\beta^*) -  Q_{nh}(\hat{\beta}_h) \leq p(\beta^*) - p(\hat{\beta}_h) - \epsilon \alpha_h^{(k)} n^{2/3} <0 
	\end{align*}
	when $n$ is large enough. Therefore the maximizer of $Q_{nh}$ must be within an $O_p(n^{-1/6})$-neighborhood of $\hat{\beta}_h$.  
	
	(2) Next, we show the consistency of $\bm{\alpha}^{(k+1)}$, $\bm{\theta}_1^{(k+1)}$,$\bm{\theta}_2^{(k+1)}$ and $\bm{\gamma}^{(k+1)}$, i.e., 
	\begin{align*}
	\bm{\alpha}^{(k+1)} - \bm{\alpha}_0 = o_p(1), ~\bm{\theta}_1^{(k+1)} - \bm{\theta}_0 = o_p(1), ~\bm{\theta}_2^{(k+1)} - \bm{\theta}_0 = o_p(1),~\bm{\gamma}^{(k+1)} - \bm{\gamma}_0 = o_p(1).
	\end{align*}
	For $l =1,\ldots,m_0$, define a new mixture distribution such that its $l$-th support points have been updated by the EM step while the rest points keep the same with $\Psi^{(k)})$, i.e., 
	\begin{align*}
	\Psi_l^{(k+1)}(\theta) & = \sum_{1\leq h \leq m_0; h\neq l} \alpha_h^{(k)} 
	\left\{\beta_h^{(k)} I(\text{P}\bm{\theta}_{1h}^{(k)} \leq \theta ) + ( 1- \beta_h^{(k)}) I(\text{P}\bm{\theta}_{2h}^{(k)} \leq \theta )  \right\}  \nn \\
	& ~~+ \alpha_l^{(k)} 
	\left\{\beta_l^{(k)} I(\text{P}\bm{\theta}_{1l}^{(k+1)} \leq \theta ) + ( 1- \beta_l^{(k)}) I(\text{P}\bm{\theta}_{2l}^{(k+1)} \leq \theta )  \right\}.
	\end{align*}
	Because EM increases the likelihood \citep{Wu1981}, we have
	\begin{align*}
	&l_n(\Psi_{l}^{(k+1)})  \geq pl_n(\Psi_{l}^{(k+1)}) \geq pl_n(\Psi_{l}^{(k)}) \geq pl_n(\hat{\Psi}_0) 
	  = l_n(\hat{\Psi}_0) + p(\bm{\beta}_0)    \geq l_n(\Psi_0) + p(\bm{\beta}_0).
	\end{align*}
	This implies $l_n(\Psi_{l}^{(k+1)}) \geq l_n(\Psi_0) $. Similarly with the proof of Lemma \ref{lm1}, we have the consistency of $\Psi_l^{(k+1)}$ for $\Psi_0$. This implies the consistency of $\bm{\theta}_{1h}^{(k+1)}$ and $\bm{\theta}_{2h}^{(k+1)}$ are consistent for $\bm{\theta}_{0l}$. This holds for every $l=1,\ldots,m_0$. Hence $\bm{\theta}_1$ and $\bm{\theta}_2$ are consistent. Similar idea can be applied on $\alpha^{(k+1)}$,$\bm{\gamma}^{(k+1)}$ as well to obtain their consistency.
	
	(3) For the last part, we derive the convergence rate of $\bm{\alpha}^{(k+1)}$, $\bm{\theta}_1^{(k+1)}$, $\bm{\theta}_2^{(k+1)}$ and $\bm{m}_1^{(k+1)}$ based on consistency result in part (2). This proceeds in the same way with the proof of Lemma \ref{lm2}. Let $
	\{l_n(\Psi^{(k+1)}(\bm{\beta}_0)) - l_n(\Psi_0)\} =\sum_{i=1}^n \log(1+\delta_i)
$,
	where
	\begin{align*}
	\delta_i & = \frac{f(D_i; \Psi^{(k+1)}(\bm{\beta}_0)) - f(D_i;\Psi_0)}{ f(D_i; \Psi_0)} \nn \\
	& = \sum_{h=1}^{m_0-1} (\alpha_h^{(k+1)}-\alpha_{0h}) \Delta_{ih} + \sum_{h=1}^{m_0} \left\{\alpha_h^{(k+1)} \{\bm{m}_{1h}^{(k+1)}\}^T \bm{Y}_i(\bm{\theta}_{0h})  +  \alpha_h^{(k+1)} \{\bm{m}_{2h}^{(k+1)} \}^T  \bm{Z}_i(\bm{\theta}_{0h}) \right\} + \sum_{i=1}^{m_0} \epsilon_{ih}
	\end{align*}
	with 
	\begin{align*}
	\epsilon_{ih} = \alpha_h^{(k+1)}  \left\{ \beta_h^{(k+1)}  \{(\bm{\theta}_{1h}^{(k+1)} -\bm{\theta}_{0h})^3\}^T \bm{U}_{ih}(\bm{\theta}_{1h}^{(k+1)})    + (1- \beta_h^{(k+1)})  \{(\bm{\theta}_{2h}^{(k+1)} - \bm{\theta}_{0h})^3\}^T \bm{U}_{ih}(\bm{\theta}_{2h}^{(k+1)})   \right\}
	\end{align*}
	and the rest quantities defined in the same way before. Let $\bar{\bm{t}}$ be 
	\begin{align*}
	\left(\alpha_1^{(k+1)} - \alpha_{01},\ldots,\alpha_{m_0-1} \{\bm{m}_{11}^{(k+1)}\}^T,\ldots, \alpha_{m_0}^{(k+1)} \{\bm{m}_{1m_0}^{(k+1)}\}^T; \alpha_1^{(k+1)} \{\bm{m}_{21}^{(k+1)}\}^T,\ldots, \alpha_{m_0}^{(k+1)} \{\bm{m}_{2m_0}^{(k+1)}\}^T\right)^T
	\end{align*}
	Then $\sum_{i=1}^n \delta_i = \sum_{i=1}^n \bar{\bm{t}}^T \bm{b}_i + \epsilon_n$ with $\epsilon_n = o_p(1) + o_p(n) \bar{\bm{t}}^T	\bar{\bm{t}}$. By the same arguments, we have $
	0 \leq 2 \bar{\bm{t}}^T\sum_{i=1}^n \bm{b}_i -n \bar{\bm{t}}^T \bm{B} \bar{\bm{t}} \{1 + o_p(1)\} + o_p(1).
	$
	Therefore $\alpha_{k+1}^{(s)} - \alpha_{0s} = O_p(n^{-1/2})$ for $s=1,\ldots,m_0-1$, $\bm{m}_{1t}^{(k+1)} = O_p(n^{-1/2})$ and $\bm{m}_{2t}^{(k+1)} = O_p(n^{-1/2})$ for every $t=1,\ldots,m_0$. This completes the proof of lemma.  
 
  \bibliographystyle{biom} 
\bibliography{ref_mixt}